\documentclass[lettersize,journal]{IEEEtran}
\usepackage{amsmath,amsfonts}
\usepackage{algorithmic}
\usepackage{algorithm}
\usepackage{array}
\usepackage[caption=false,font=normalsize,labelfont=sf,textfont=sf]{subfig}
\usepackage{textcomp}
\usepackage{stfloats}
\usepackage{url}
\usepackage{verbatim}
\usepackage{graphicx}
\usepackage{multirow}
\usepackage{tabularray}
\usepackage{cite}

\begin{document}

\title{Personalized Emotion Detection from Floor Vibrations Induced by Footsteps}

\author{Yuyan Wu, Yiwen Dong, Sumer Vaid, Gabriella M. Harari, Hae Young Noh
\thanks{Yuyan Wu, Yiwen Dong and Hae Young Noh are with the Department of Civil and Environmental Engineering, Stanford University, Stanford, CA, USA (e-mail: wuyuyan@stanford.edu; ywdong@stanford.edu; noh@stanford.edu).}
\thanks{Sumer Vaid is with Harvard Business School, Harvard University, Boston, MA, USA (e-mail: svaid@hbs.edu).}
\thanks{Gabriella M. Harari is with the Department of Communication, Stanford University, Stanford, CA, USA (email: gharari@stanford.edu).}
}


\IEEEpubid{}

\maketitle

\begin{abstract}
Emotion recognition is critical for various applications, including the early detection of mental health disorders and emotion-based smart home systems. Previous studies utilized various sensing methods for emotion recognition, such as wearable sensors, cameras, and microphones. However, these methods have limitations in long-term domestic use because of the inherent limitations, including intrusiveness and privacy concerns.
To overcome these limitations, this paper introduces a non-intrusive and privacy-friendly personalized emotion recognition system, EmotionVibe, which leverages footstep-induced floor vibrations for emotion recognition. The main idea of EmotionVibe is that individuals’ emotional states influence their gait patterns, subsequently affecting the floor vibrations induced by their footsteps. However, there are two main research challenges: 1) the complex and indirect relationship between human emotions and footstep-induced floor vibrations and 2) the large between-person variations within the relationship between emotions and gait patterns. To address these challenges, we first empirically characterize this complex relationship and develop an emotion-sensitive feature set including gait-related and vibration-related features from footstep-induced floor vibrations. Furthermore, we personalize the emotion recognition system for each user by calculating gait similarities between the target person (i.e., the person whose emotions we aim to recognize) and those in the training dataset and assigning greater weights to training people with similar gait patterns in the loss function. 
We evaluated our system in a real-world walking experiment with 20 participants, summing up to 37,001 footstep samples. EmotionVibe achieved the mean absolute error (MAE) of 1.11 and 1.07 for valence (unpleasant to pleasant) and arousal (calm to excited) score estimations, respectively, reflecting 19.0$\%$ and 25.7$\%$ error reduction compared to the baseline method.
\end{abstract}

\begin{IEEEkeywords}
Emotion Recognition, Footstep-Induced Floor Vibrations, Mental Health Monitoring
\end{IEEEkeywords}

\section{Introduction}
\label{sec:intro}

Emotion recognition is crucial for various applications, such as mental health monitoring and emotion-based smart home devices~\cite{singh2023emotion, rajuroyenergy, fodor2023real}. According to the National Institute of Mental Health (NIMH), 23.1$\%$ U.S. adults, approximately 59.3 million individuals live with a mental illness in 2022~\cite{NIMH2023MentalIllness}. On average, individuals with anxiety or depression have a lifespan of 7.9 years shorter than those without these conditions~\cite{pratt2016excess}. Since mental health disorders are often characterized by increased emotional instability and fluctuations~\cite{nelson2020everyday}, monitoring changes in individuals' emotions can facilitate the early detection and intervention of severe mental illnesses~\cite{lee2019early}. 
In addition, integrating emotion recognition systems into smart home devices enables adaptive recommendation systems that enhance user interaction. For example, emotional data can inform music recommendations, adjust lighting and temperature for comfort or relaxation~\cite{kaminskas2012contextual, mostafavi2024effects}, and provide personalized content suggestions for smart TVs and video games~\cite{lee2014relationship, frommel2018towards}.

In previous work, various sensing methods have been used for emotion recognition, including wearable sensors~\cite{shu2018review,raheel2020physiological}, cameras~\cite{ko2018brief, huang2023study}, microphones~\cite{popova2018emotion, ooi2014new}, and multi-modal sensing methods~\cite{ahmed2023systematic, zhang2023deep}. However, these methods are limited in long-term domestic usage because of intrusiveness or privacy concerns. Wearable sensors are intrusive and may cause discomfort. Cameras and microphones, which rely on capturing facial expressions, body postures, or speaking voices, often raise privacy concerns. Multi-modal sensing requires increased hardware deployment and data processing complexity. These limitations restrict the widespread use of emotion recognition systems for indoor applications.

\begin{figure}[!t]
  \centering
  \includegraphics[width=0.9\linewidth]{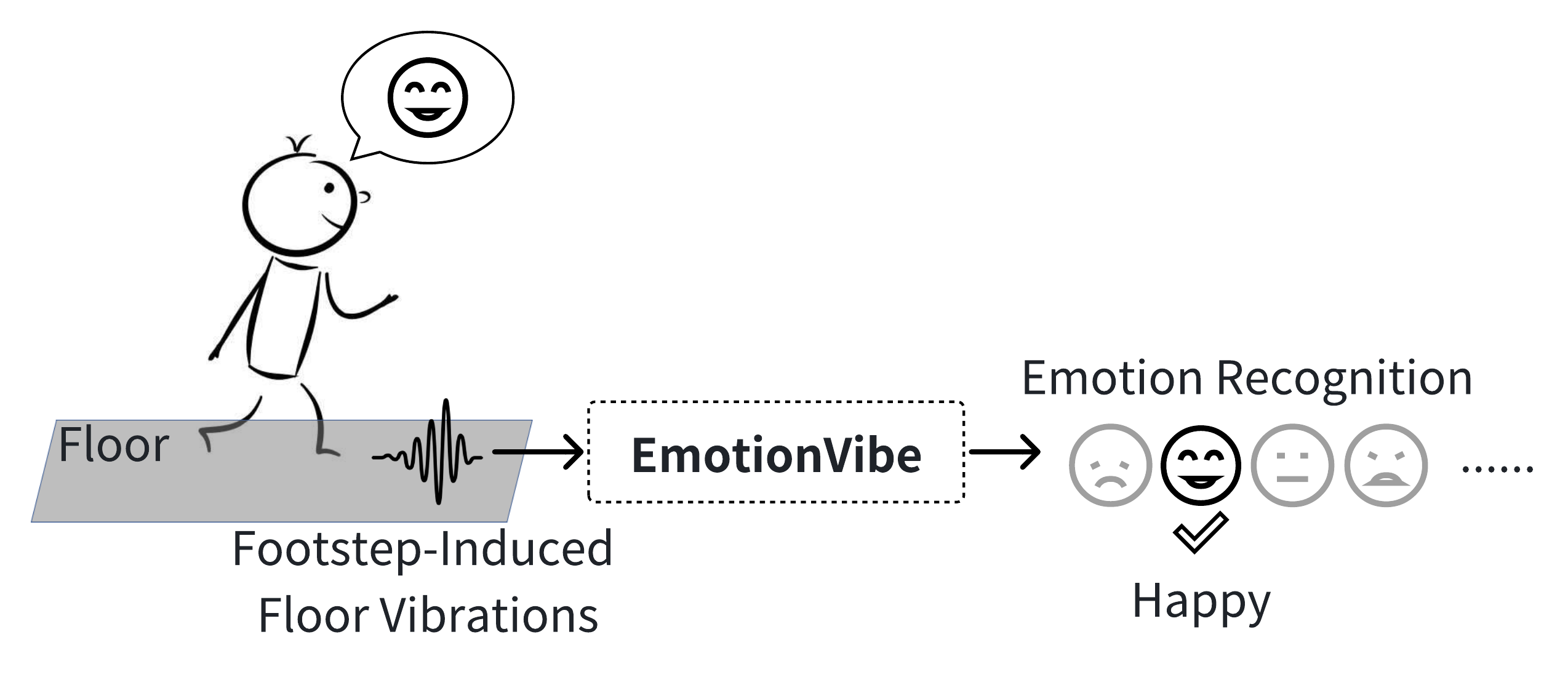}
  \caption{EmotionVibe recognizes human emotions by analyzing the footstep-induced floor vibration patterns, which are affected by emotions.}
  \label{fig:intro_fig}
\end{figure}

This paper introduces EmotionVibe, a novel personalized emotion recognition system that infers human emotions through footstep-induced floor vibrations (see Fig.~\ref{fig:intro_fig}). The main intuition of EmotionVibe is that people's emotion influences their gait patterns, which in turn affect the floor vibrations induced by their footsteps. To this end, by analyzing footstep-induced floor vibration patterns captured by the vibration sensors attached to the floor, the emotional state of the pedestrian can be inferred. We showed a proof of concept for this intuition through preliminary small-scale laboratory experiments in our previous work~\cite{wu2023emotion}. Compared to other sensing methods, EmotionVibe offers a non-intrusive and privacy-friendly personalized emotion recognition method.

However, recognizing emotions through footstep-induced floor vibrations is challenging due to the complex, indirect, and personalized relationship between human emotions and footstep-induced floor vibrations. The main challenges are:

1) \textit{The complex and indirect relationship between human emotions and footstep-induced floor vibrations.} Emotions influence human gait patterns in various aspects, including kinematics, kinetics, and spatiotemporal parameters~\cite{homagain2023emotional, roether2009critical, li2016emotion, xu2022emotion, montepare1987identification}.
Furthermore, the relationship between gait patterns and footstep-induced floor vibrations is also challenging to model with the complex mechanisms of foot-floor interaction~\cite{dong2025modeling}. 

2)
\textit{Large between-person variations in the relationship between human emotions and footstep-induced floor vibrations.} Each person has a distinct walking style and responds uniquely under varying emotional conditions~\cite{pan2017footprintid, dong2024home}. These variations result in highly diverse gait characteristics and footstep-induced floor vibration patterns. Consequently, vibration data from people whose gait patterns significantly differ from the target person can reduce the performance of the emotion recognition model.

To capture the complex relationship between human emotions and footstep-induced floor vibrations, we develop emotion-sensitive features from two aspects: gait-related and vibration-related features. The selection of these features is inspired by an analysis of how emotions influence gait pattern parameters including kinematic, kinetic, and spatiotemporal parameters, and how these gait patterns impact the characteristics of footstep-induced floor vibrations.

To address the large between-person variations in the relationship between emotions and gait patterns, we personalize the emotion recognition model by assigning higher weights to training samples from people whose gait patterns are similar to the target person in the loss function. To achieve this, we first estimate the gait similarity indices between the target person and people in the training dataset based on the distance between the features embedded in a lower dimensional space representing their gait patterns. We then utilize these gait similarity indices as sample weights in the loss function to personalize the emotion recognition model. Consequently, the performance of the emotion recognition model improves for the target person, as it assigns greater importance to the training data that are more similar to the target person.

The main contributions of this work are:
\begin{itemize}
\item We introduce EmotionVibe, a novel personalized emotion recognition system based on footstep-induced floor vibrations. EmotionVibe provides a non-intrusive and privacy-friendly approach to emotion recognition, making it well-suited for in-home applications.

\item We develop two sets of emotion-sensitive features to capture the complex relationship between emotion and footstep-induced floor vibrations. In addition, we personalize the emotion recognition model for each target user based on the gait similarity between people.

\item We evaluated EmotionVibe in real-world experiments with 20 participants, achieving promising results for emotion score estimation.
\end{itemize}

The rest of the paper is organized as follows: Section~\ref{sec:related_work} discusses related works and provides a comparison between the related works and our system. Section~\ref{sec:relation} introduces the basic physical insights of our system, including the description of emotion, the effect of emotions on gait patterns, and the effect of gait patterns on footstep-induced floor vibrations. Section~\ref{sec:method} details the EmotionVibe system design. Section~\ref{sec:eva} shows the real-world experiments and the evaluation results of EmotionVibe. Section~\ref{sec:discussion} evaluates the effectiveness of each module in EmotionVibe, assesses the robustness of our system performance, and explains the rationale behind participant selection criteria. Finally, Section~\ref{sec:conclusion} concludes our work and explores future directions.
\section{Related Works}
\label{sec:related_work}
This section provides an overview of current research on various emotion recognition sensing methods as well as vibration analysis and modeling methods.

\subsection{Emotion Recognition Methods}

Previous studies on emotion recognition can be divided into: physiological signal-based methods, facial indicator-based methods, body behavior-based methods, and linguistic-based methods.

Physiological signals such as Electroencephalography (EEG)~\cite{liu2021review, bos2006eeg,petrantonakis2009emotion}, Electrocardiography (ECG)~\cite{hasnul2021electrocardiogram, dissanayake2019ensemble,jing2009research}, Galvanic Skin Response (GSR)~\cite{liapis2015recognizing,greco2016skin}, Heart Rate Variability (HRV)~\cite{quintana2012heart, appelhans2006heart} are effective indicators of human emotions. These physiological signals are difficult to mimic and provide accurate results in emotion recognition. However, measuring these physiological signals typically requires direct sensor-skin contact, which can cause discomfort and limit daily usage. Moreover, their effectiveness is often reduced in mobile settings due to motion artifacts and signal instability.

Other methods use visual cues from the face, including expressions~\cite{ioannou2005emotion, tarnowski2017emotion, liu2017facial}, eye movements~\cite{liapis2015recognizing, yan2021simplifying}, and gaze patterns~\cite{bal2010emotion, wieckowski2017eye}. These methods are usually camera-based, thus enabling contactless emotion recognition and overcoming the limitations of wearable sensors. However, these methods are limited by lighting conditions, visual obstacles, and camera shooting angles. In addition, cameras usually raise privacy concerns, thus limiting their application in domestic settings.

Body behavior is another important indicator of human emotions. Previous research has utilized cameras~\cite{mingming2020emotion, wang2015adaptive}, motion capture systems~\cite{bhatia2022motion, kapur2005gesture}, smartwatches~\cite{quiroz2017emotion, sultana2020using}, smartphones~\cite{kolakowska2020review, olsen2016smartphone, piskioulis2021emotion}, and force platforms~\cite{janssen2008recognition} to analyze human gait behavior, posture, and other body movements to infer the user's emotional state.
Camera and motion-capturing-based methods are contactless methods that capture a broad range of movements and can extract skeletal and joint locations for emotion recognition~\cite{ouguz2024emotion, daoudi2017emotion,barliya2013expression,li2016identifying,karg2010recognition}. Compared to facial expression recognition-based emotion recognition, these methods are effective when the subject is away from the camera or when facial features are obscured. However, they can still cause privacy concerns for in-home scenarios and are limited by visual obstacles. The wearable or mobile device-based methods, use body-attached or embedded sensors in mobile devices to record movements directly for emotion recognition ~\cite{quiroz2018emotion,quiroz2017emotion,zhang2016emotion,cui2016emotion,liu2019automatical}. These devices, equipped with accelerometers and gyroscopes, realize emotion recognition by analyzing the occupants' movement information. Although more convenient than wired sensors, they still require users to carry or wear the device which limits their everyday applicability. The force-based methods, involve measuring ground reaction forces using force platforms during walking to recognize emotions~\cite{janssen2008recognition}.  However, the limited coverage of force platforms necessitates dense deployment, restricting their suitability for in-home use.

Emotion recognition through linguistic analysis mainly includes speech-based and text-based methods. Speech-based methods analyze vocal attributes such as pitch, tone, speech rate, and intensity, and variations in these features for emotion recognition~\cite{schelinski2019relation, larrouy2024sound}. For instance, a higher pitch and faster speech indicate excitement, while a slower rate and softer tone suggest sadness.
Text-based methods use Natural Language Processing (NLP) to analyze word choice, sentence structure, and contextual semantics for emotion detection. Advances in Large Language Models have enhanced sensitivity to linguistic nuances and context~\cite{wang2024large, pico2024exploring}. However, speech-based methods require high-quality audio, limiting effectiveness in uncontrolled environments and raising privacy concerns due to microphone usage. Text-based methods depend on user-provided textual data, which may not always be available.

Compared to existing methods, EmotionVibe provides a non-intrusive and privacy-friendly approach to emotion recognition by utilizing footstep-induced floor vibrations. Unlike physiological, facial, or body behavior-based techniques that require wearables or cameras, EmotionVibe captures emotional cues without direct skin contact or compromising privacy. Moreover, it does not rely on verbal or text input from users. Consequently, EmotionVibe presents a promising solution for real-world, long-term domestic applications.

\subsection{Vibration Analysis and Modeling}

In this subsection, we review papers related to vibration and audio signal processing, modeling, and effective machine-learning model simplification methods that inspire the development of emotion-sensitive feature sets and the emotion recognition model.

Previous studies proposed various feature extraction methods for vibration and audio signal processing to capture signal features, which can be categorized into classical and deep learning-based methods. Classical methods extract features based on interpretable physical principles of signal processing research, including time-domain features (mean, kurtosis, zero-crossing rate, envelope analysis), frequency-domain features (Fourier transform, power spectral density), and time-frequency-domain spectral features (mel-frequency cepstral coefficients, short-time Fourier transform, constant Q transform, and wavelet transform)~\cite{sharma2020trends, lartillot2007matlab,ding2022feature, yang2003vibration,zhang2022vibration}. These features can handle transient and non-stationary signals and are used for pattern analysis in speech recognition~\cite{yu2016automatic}, speaker identification~\cite{campbell1997speaker}, structural health monitoring~\cite{zhang2022vibration}, and human-building interaction studies~\cite{ellis1997human}. These classical features are interpretable and computationally efficient but often rely on domain expertise to design effective feature sets. On the other hand, deep learning-based methods, including convolutional neural networks and autoencoders, can automatically extract complex hierarchical features from raw signals or spectrograms, enabling robust performance in tasks such as fault detection and sound classification~\cite{bose2020deep, zhang2022vibration, wang2018automated, lin2017structural}. However, they are computationally intensive and require a large number of high-quality datasets for learning the implicit relationships. EmotionVibe is a combination of these two approaches. We first analyze the complex relationship between emotion and footstep-induced floor vibrations and develop an emotion-sensitive feature set based on this relationship, including gait-related and vibration-related features. Then, we model these features using a deep learning framework to learn implicit relationships that cannot be directly given by classical features.

Due to the limited dataset resulting from the high cost of human experiments, EmotionVibe employs an iterative pruning approach, thereby mitigating overfitting and enhancing model effectiveness in emotion recognition. Previous model simplification methods mainly consist of pruning and knowledge distillation. Pruning methods remove unnecessary parameters or structures in the model to reduce the model size and computational cost. It can be categorized into unstructured pruning that removes individual weights~\cite{han2015learning, frankle2018lottery} and structured pruning that prunes network layers~\cite{anwar2017structured, lemaire2019structured, el2022data}. Unstructured pruning targets individual weights and can remove redundant parameters in a more fine-grained manner. However, the inference speed for the unstructured pruned models is limited by the sparse weight matrices caused by such fine-grained pruning. In contrast, structured pruning supports faster inference. However, it is coarse-grained pruning, with each pruning cutting out an entire unit. This may cause important information loss~\cite{yang2021comparative}. The knowledge distillation method transfers knowledge from a larger, pre-trained teacher model to a smaller student model, effectively simplifying the model while maintaining performance~\cite{gou2021knowledge, cho2019efficacy, park2019relational}. However, this method first requires a well-trained teacher model, which relies on a large dataset. This does not apply to our study because of the limited data we collected in the laboratory experiments. EmotionVibe simplifies the network structure by iteratively cutting the least important parameters in the network during training to reduce overfitting. Since reducing computation time is not our top priority, we chose the unstructured pruning method, which is more fine-grained and has a lower probability of cutting important parameters.

\section{Characterizing the Relationship Between Emotions and Footstep-Induced Floor Vibrations}
\label{sec:relation}
\begin{figure*}[ht] 
  \centering
  \includegraphics[width=0.85\textwidth]{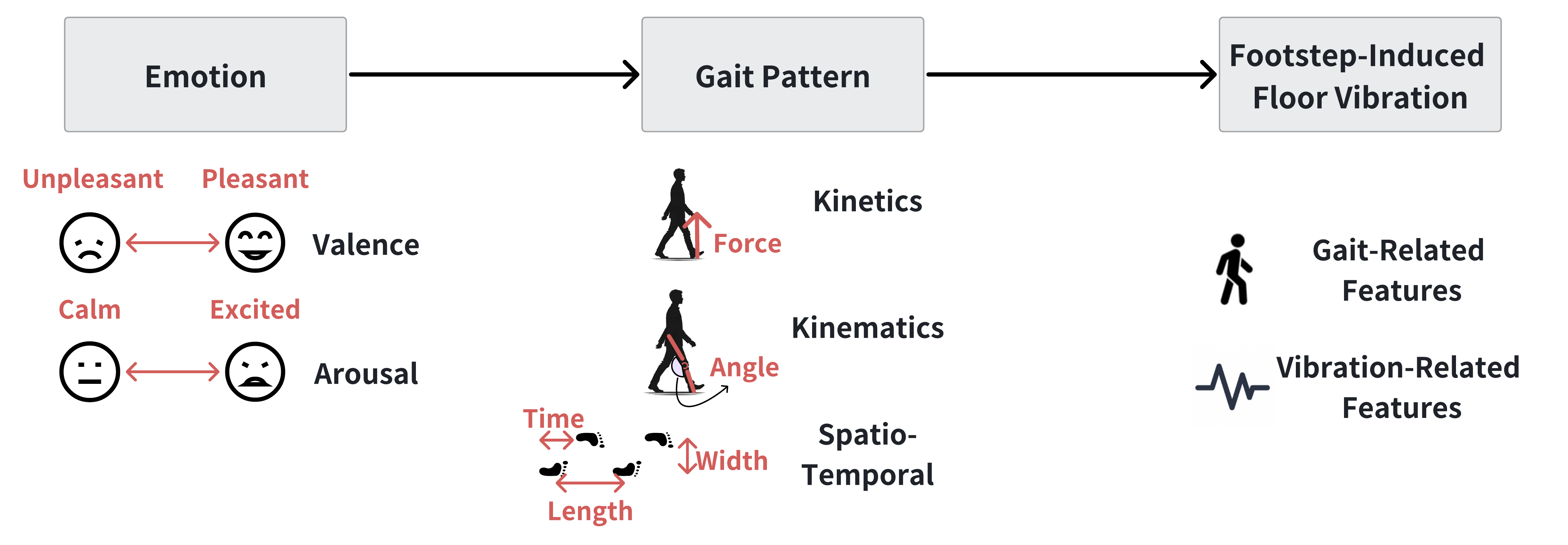} 
  \caption{Main Intuition of EmotionVibe. Emotion impacts gait pattern, which in turn affects footstep-induced floor vibrations.}
  \label{fig:background}
\end{figure*}

EmotionVibe is based on the insight that human emotions affect their gait patterns, subsequently influencing the footstep-induced floor vibration patterns (see Fig.~\ref{fig:background}). To this end, we can infer people's emotional states by analyzing the footstep-induced floor vibration patterns captured by the vibration sensors attached to the floor. In this section, we analyze and characterize the relationship between human emotions and footstep-induced floor vibrations through both analytical and empirical studies.

\subsection{Description of Emotions}\label{sec:des_emotion}

\begin{figure}[!t]
  \centering
  \includegraphics[width=0.9\linewidth]{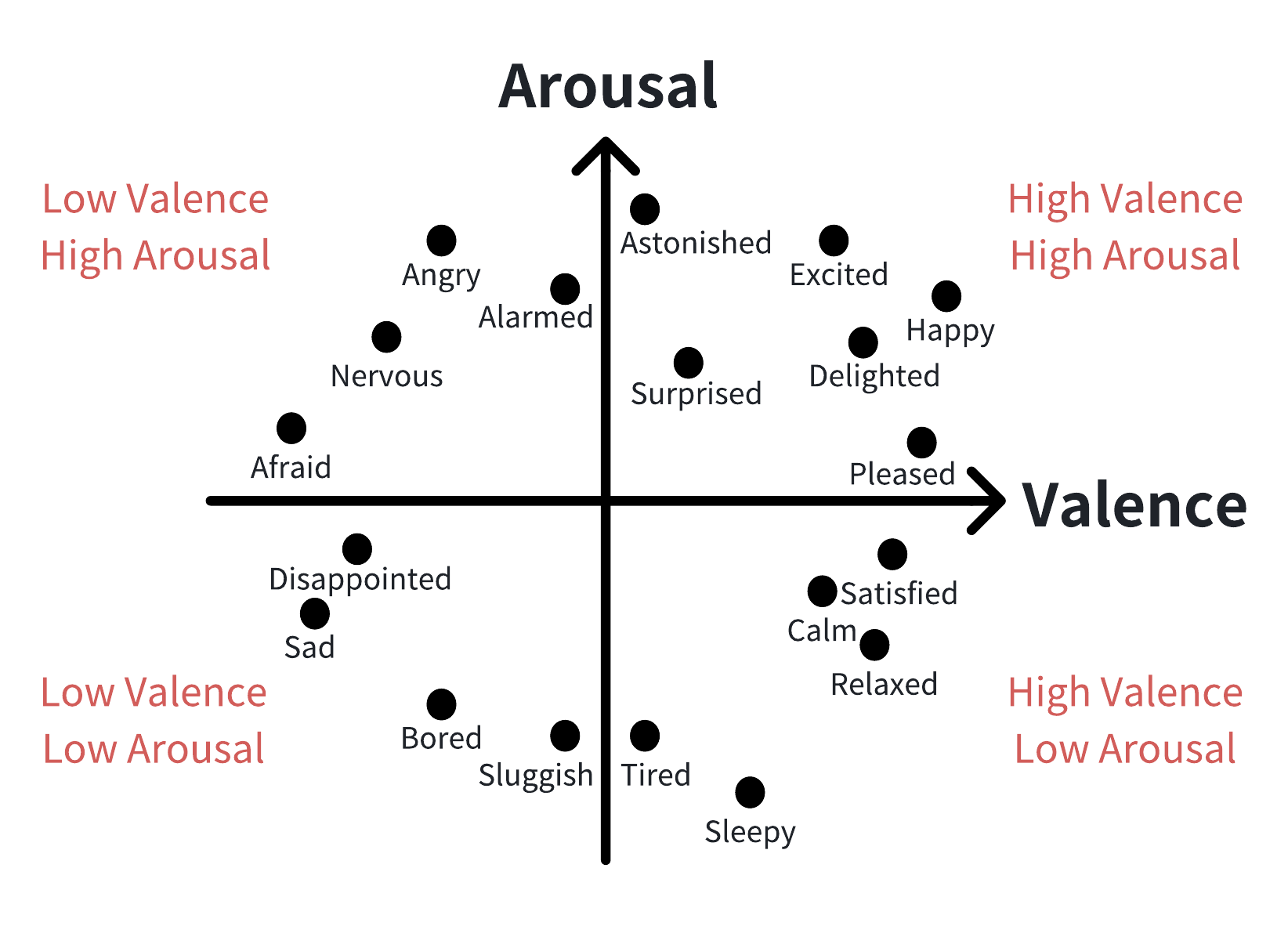}
  \caption{Description of Emotion Using Valence-Arousal Model~\cite{russell1980circumplex}.}
  \label{fig:emotion_model}
\end{figure}

In the field of psychology, emotions are often described by the two-dimensional valence-arousal model, as shown in Fig.~\ref{fig:emotion_model}~\cite{russell1980circumplex}. Emotions map onto a 2D space formed by the valence and the arousal axes. Valence describes the extent to which an emotion is positive or negative.  In other words, it captures the degree to which an emotion is pleasant or unpleasant~\cite{frijda1986emotions}. High-valence emotions include happiness, joy, or excitement, while low-valence emotions include sadness, anger, or fear. Arousal refers to the physiological and psychological state of being awake or reactive to stimuli. It ranges from calmness and sleepiness at the low end to increased excitement and heightened alertness at the high end. High arousal is characterized by emotions of being energized, alert, or excited, while low arousal is associated with emotional states such as calmness, relaxation, or lethargy. These two dimensions are assumed to be independent of each other. Based on this 2D model, human emotions can be basically classified into four classes: high-valence, high-arousal (HVHA), high-valence, low-arousal (HVLA), low-valence, high-arousal (LVHA), and low-valence, low-arousal (LVLA). Each type of emotion maps to a specific area in this 2D space. For instance, excitement is a high-valence, high-arousal emotion; anger is a low-valence, high-arousal emotion; relaxation is a high-valence, low-arousal emotion; and depression is a low-valence, low-arousal emotion. 

\subsection{Effect of Emotion on Gait Patterns}

Emotions impact gait patterns from various aspects, including kinetics, kinematics, and spatiotemporal gait parameters (see Fig.~\ref{fig:background})~\cite{homagain2023emotional, roether2009critical, li2016emotion, xu2022emotion, montepare1987identification}. Kinetics represents the forces and moments that cause the motion. This includes the ground reaction forces, joint forces, and moments that influence the movement of limbs. Kinematics focuses on the geometric movement patterns of the body without considering the forces related to these movements. For example, it involves how the angle of the knee changes throughout a stride, the trajectory of the foot during the swing phase, and the range of rotation in the hip joint. Spatial-temporal gait parameters measure time-related and distance-related gait characteristics, such as stride length, step length, stride width, gait speed, cadence, stance time, and swing time.

Emotions affect the kinetic aspects of gait, i.e., the forces that cause or result from motion, by influencing the force amplitudes and pressure distributions applied during walking~\cite{gross2012effort,xu2003emotion,janssen2008recognition}.
High-arousal emotions, such as anger and happiness, are associated with heightened forcefulness and greater stride intensity, whereas low-arousal emotions like sadness are related to a less energetic gait~\cite{gross2012effort,xu2003emotion,roether2009critical}. In addition, variations in plantar pressure distribution and center of pressure shifts across emotional states reflect changes in foot loading and balance~\cite{janssen2008recognition}.

 The kinematics of locomotion are another important indicator of emotions in gaits~\cite{roether2009critical, xu2003emotion,feldman2020gait,michalak2009embodiment,gross2012effort,barliya2013expression}. Kinematics of locomotion mainly include the movements of joint angles and body parts. It is a description of geometric motion without considering the footstep forces. Sadness is characterized by reduced amplitudes of pelvic rotation, hip flexion, shoulder flexion, torso rotation, and elbow flexion, leading to a more slumped posture and decreased arm swing amplitude~\cite{roether2009critical, xu2003emotion,feldman2020gait,michalak2009embodiment}. 
Angry individuals exhibit more flexed trunks, elevated shoulders, and a forward-inclined spine, along with increased amplitudes in shoulder, elbow, hip, and knee movements, indicating a more dynamic and aggressive gait pattern compared to neutral or sad states ~\cite{roether2009critical,gross2012effort}.
Joyful and content lead to an upright torso posture, with increased neck, trunk, thigh elevation angles, and thoracic extension, and elevated amplitudes in shoulder, elbow, trunk, pelvis, and hip movements, suggesting a more open, energetic, and coordinated gait compared to sadness~\cite{roether2009critical,barliya2013expression}.

Furthermore, spatiotemporal gait parameters, which refer to time-related and distance-related gait characteristics described by gait parameters, are also influenced by human emotions~\cite{naugle2011emotional,kang2016effect,xu2003emotion,du2023spatiotemporal,lemke2000spatiotemporal,homagain2023emotional,gross2012effort,feldman2020gait}. These parameters include gait velocity, stride, step lengths, step width, durations of single and double support, swing periods, phases, step frequency, and other parameters within a gait cycle. 
For example, happiness leads to increased stride length, pace, and walking speed, suggesting a more energetic state~\cite{naugle2011emotional,kang2016effect,xu2003emotion}.
Sadness reduces gait velocity and arm movement, shortens stride length, and increases double limb support, cycle duration, step time, stance time, and swing time ~\cite{du2023spatiotemporal,lemke2000spatiotemporal,homagain2023emotional,feldman2020gait}.
Anger usually results in faster walking speed and larger stride lengths, reflecting increased movement energy and expansiveness in body language~\cite{gross2012effort,kang2016effect}.
In addition, fear and excitement lead to reduced step time compared to neutral conditions~\cite{lemke2000spatiotemporal}.

\subsection{Effect of Gaits on Footstep-Induced Floor Vibrations}

Previous works have shown that footstep-induced floor vibrations contain valuable information about pedestrians' gait patterns~\cite{dong2020md,pan2017footprintid,dong2023characterizing,dong2023detecting,mirshekari2018occupant,fagert2019gait, wu2023emotion,drira2019occupant,drira2021using}. Footstep-induced floor vibration signals can be used for person identification~\cite{pan2017footprintid,dong2023characterizing,pan2015indoor,drira2019occupant,drira2021using}, gait balance detection~\cite{dong2020md,fagert2017characterizing}, and inference of gait parameters, foot-floor contact types, and ground reaction force~\cite{dong2024ubiquitous,dong2023detecting,fagert2021structure}. These findings provide a foundation for exploring the effects of emotional states on gait changes. The analysis of footstep-induced floor vibrations mainly incorporates time-domain and frequency-domain features, along with other task-specific characteristics related to the vibration signals.
The spatio-temporal gait parameters directly affect the temporal and spectral components of the floor vibrations. For example, step frequency affects the time difference between footstep vibration pulses, and stride length affects the energy difference between adjacent footstep vibration signals in the vibration signal. The gait kinetic information affects the energy profile of the floor vibration.

\begin{figure}[!t]
  \centering
  \includegraphics[width=\linewidth]{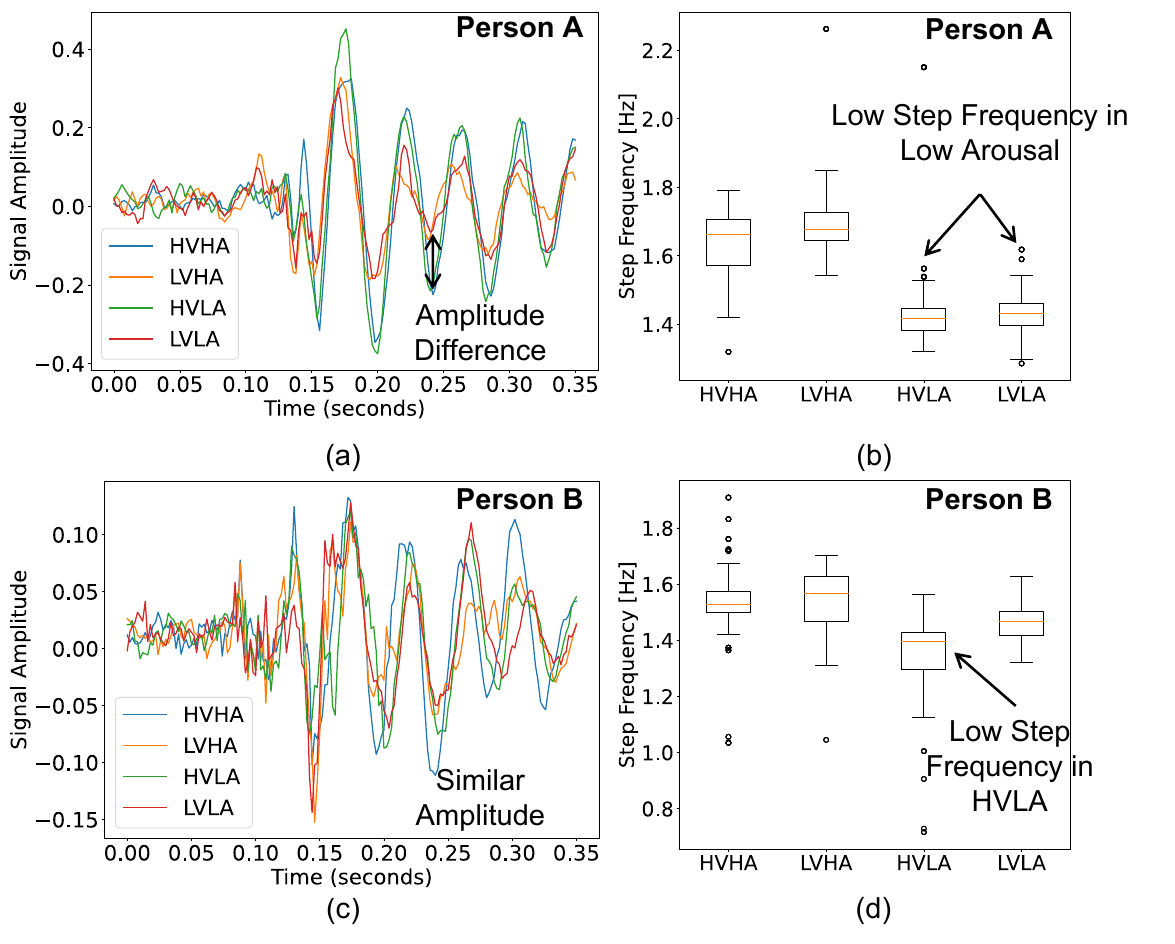}
  \caption{Representative example for footstep-induced floor vibration patterns characterization under different emotional states\protect\footnotemark: (a, c) Average vibration signals for Person A and Person B across various emotional states. Vibration amplitudes are higher during emotions with high valence for person A. (b, d) Variations in step frequency for Person A and Person B under different emotional states. Step frequency is larger during high arousal emotional states for person A and lower during high valence low arousal states for person B. Both individuals exhibit distinct footstep-induced vibration patterns across different emotional states. In addition, individual variability is not negligible.}
  \label{fig:characterization}
\end{figure}

We characterize the effect of emotional state on the footstep-induced floor vibration patterns based on empirical data collected in laboratory experiments. For Person A, the footstep amplitude is significantly larger during states of high valence emotions, and the step frequency notably increases in high arousal situations (see Fig.~\ref{fig:characterization} (a, b)). These observations are consistent with previous conclusions in the literature that high arousal emotions are always associated with faster walking speeds, and that happiness elicits more forceful footsteps~\cite{naugle2011emotional,kang2016effect,xu2003emotion}. However, this pattern does not apply to Person B (see Fig.~\ref{fig:characterization} (c, d)). Relying solely on features identified in previous studies would yield inadequate emotion recognition results, as they fail to capture the complex relationship between human emotions and footstep-induced vibrations. Other gait behaviors related to emotions, such as leaning backward when relaxed or reduced foot-lift height when depressed, are also important for emotion recognition. Consequently, developing a comprehensive emotion-sensitive feature set encompassing various aspects of footstep-induced vibrations is essential for accurate emotion recognition.

\footnotetext{HVHA: high-valence, high-arousal; LVHA: low-valence, high-arousal; HVLA: high-valence, low-arousal; LVLA: low-valence, low-arousal.}

\begin{figure}[!t]
  \centering
  \includegraphics[width=\linewidth]{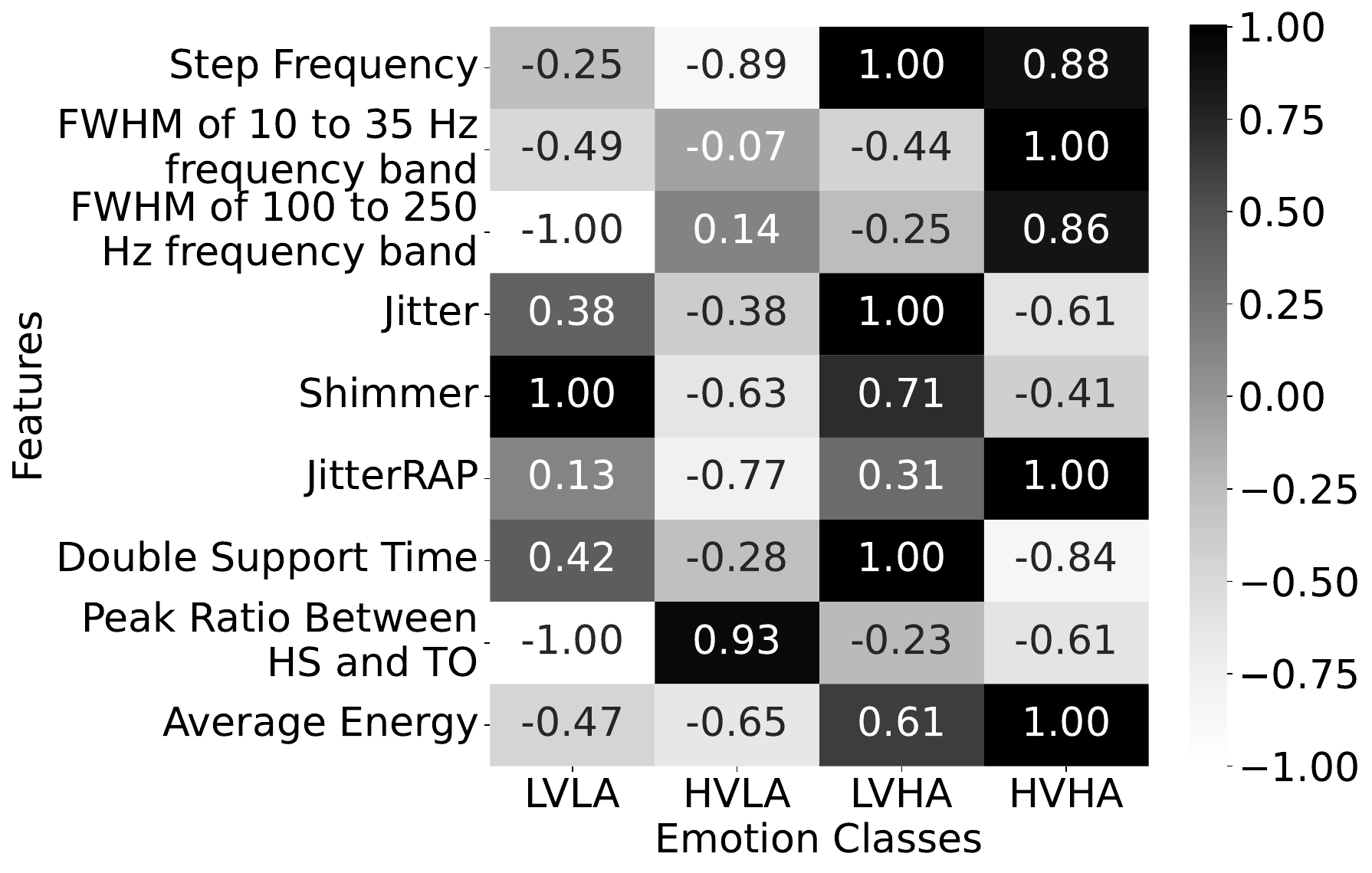}
  \caption{Feature variance across the four emotion classes. Each cell indicates the deviation from the overall average for a given feature under specific emotional conditions. Positive values (towards 1.00) suggest that the feature increases in the corresponding emotion classes, while negative values (towards -1.00) denote that the feature decreases in the corresponding emotion classes. The effectiveness of these features is evident in their varying performance across different emotion classes. }
  \label{fig:feature_color}
\end{figure}

To this end, we develop an emotion-sensitive feature set including gait-related and vibration-related features to capture the complex relationship between human emotions and floor vibrations. 
The feature set is detailed in Section~\ref{sec:feature}. Due to the large number of features, we show representative characterization examples in Fig.~\ref{fig:feature_color}.
The heatmap illustrates the deviations across the four emotion classes. The distributions of features vary depending on emotional states. For example, under high-arousal conditions, the step frequency increases, indicating faster walking; the peak height ratio between heel-strike and toe-off increases during the high valence low arousal situations, indicating a tendency to lean backward; the average energy of the footstep-induced floor vibration signals increases during high-valence emotional situations, indicating a more heavy-footed walking pattern. This finding aligns with conclusions from other research in gait emotion analysis~\cite{gross2012effort,xu2003emotion}. This variation in feature distributions shows the feasibility of using these features for emotion recognition.

Furthermore, each feature exhibits distinct efficiency in differentiating specific emotions, which shows the need to combine these features for effective emotion recognition. For example, step frequency and average energy are effective in distinguishing between high and low arousal emotions, but less effective for differentiating high and low valence. Conversely, the peak ratio between heel strike (HS) and toe-off (TO) separates high valence-low arousal emotions (HVLA) but shows minimal variation for other emotions.

\section{EmotionVibe: Emotion Recognition System Using Footstep-Induced Floor Vibrations}
\label{sec:method}

\begin{figure}[!t]
  \centering
  \includegraphics[width=\linewidth]{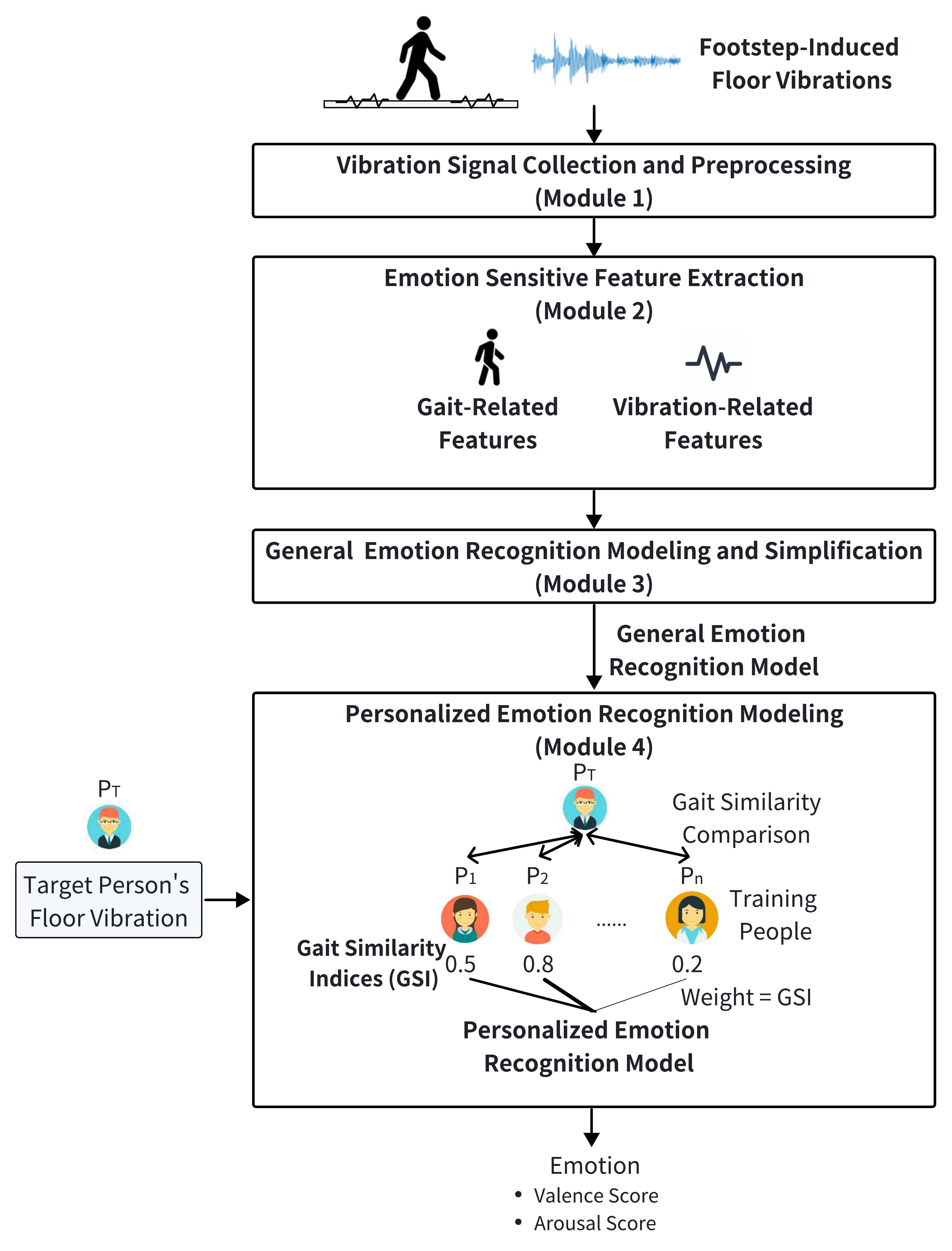}
  \caption{EmotionVibe System Overview. 
  } 
  \label{fig:system}
\end{figure}

The EmotionVibe system collects and analyzes floor vibration signals induced by footsteps to recognize the pedestrian's emotions with four modules: 1) Vibration signal collection and preprocessing, 2) Emotion-sensitive feature extraction, 3) General emotion recognition modeling and simplification, and 4) Personalized emotion recognition modeling (see Fig.~\ref{fig:system}). 
After collecting and preprocessing footstep-induced floor vibration data (Module 1), we capture the complex relationship between human emotion and floor vibrations by developing emotion-sensitive feature sets to address the first challenge (Module 2). These features, inspired by the relationship between human emotions and floor vibrations discussed in Section~\ref{sec:relation}, include gait-related and vibration-related features.
Next, a general emotion recognition model is developed for preliminary emotion recognition and data-driven emotion information extraction from features (Module 3). This model is subsequently personalized to the target person (i.e., the person whose emotions we aim to recognize) to tackle the second challenge (Module 4). The personalization is achieved by first comparing the gait similarities between the vibration data from the target person and the training people. These gait similarities are then employed as weighting factors in the loss function during model fine-tuning, resulting in a personalized emotion recognition model. Finally, EmotionVibe outputs the target person's emotion estimation results, represented by valence and arousal scores.

\subsection{Module 1: Vibration Signal Collection and Preprocessing}

The vibration signal collection and preprocessing module includes three steps: 1) footstep-induced floor vibration signal collection, 2) single footstep signal segmentation, and 3) signal clipping calibration. First, we collect vertical floor vibrations using geophone sensors. Geophones are selected for their low cost and high sensitivity to vertical floor velocity in the low-frequency range (0–200 Hz), which aligns with the range of footstep-induced floor vibrations~\cite{pan2017footprintid}. The collected vibrations are then sent to an amplifier board to enhance their signal-to-noise ratio. Next, we segment the vibration signals representing each footstep (defined as a single footstep signal) for gait analysis by identifying the prominent peak in the wavelet transform coefficients within the frequency band of dominant structural natural frequencies (selected based on the floor type and specified in Section~\ref{sec:hyper}). This peak corresponds to the impulse force induced by a single footstep. The signal window of the average footstep duration (chosen as 0.35 s based on our observation) is then extracted around this peak to segment the single footstep signal. Finally, to enhance signal quality,  we identify the clipped signal sections and then use polynomial interpolation to reconstruct the clipped signal section using the neighboring data points of the clipped sections. The clipped sections are detected when the signal reaches the upper or lower limit that the sensor can measure and remains at the limit for at least 3 samples, representing the section where the signal exceeds the range of the vibration sensor. Assuming the signal remains continuous, the signal is reconstructed by fitting a polynomial function with the neighboring points of the clipped section and then using this function to interpolate the data within the clipped sections. After interpolation, we obtain a continuous signal representation for the clipped sections with enhanced signal quality.

\subsection{Module 2: Emotion-Sensitive Feature Extraction}
\label{sec:feature}
To capture the complex relationship between human emotions and floor vibrations, we develop gait-related and vibration-related features from the preprocessed single footstep signals. Gait-related features capture gait parameters influenced by emotions during the gait cycle, reflecting footstep characteristics. Vibration-based features, on the other hand, capture the detailed characteristics of vibration signals induced by footsteps.
The selection of these features is inspired by our investigation of the relationship between human emotions and footstep-induced floor vibrations (see Section~\ref{sec:relation}). 

\subsubsection{Gait-Related Features} Gait-related features are based on gait parameters in the gait cycle and provide basic information about footstep patterns influenced by emotions. Gait-related features include a variety of gait parameters: step frequency, double support time, peak height ratio of heel-strike and toe-off, full width at half maximum (FWHM) of heel-strike and toe-off, and energy contours (including the raw, smoothed, and logarithm energy contour). 

The gait-related features are chosen based on the influence of human emotions on gait patterns and the corresponding affected gait parameters (see Section ~\ref{sec:relation}).
For example, step frequency usually increases under emotions like anxiety or happiness, while a slower step frequency might reflect sadness or fatigue~\cite{roether2009critical, homagain2023emotional}. Double support time can help identify cautious and confident walking~\cite{bansal2021does}. The peak-to-height ratio between heel-strike and toe-off indicates the angle between the foot and floor during the stance phase and is related to the degree to which a person's center of gravity is tilted back (larger during relax states when leaning backward)~\cite{kaushik2021perception}. In addition, the FWHM indicates the duration of the heel-strike and toe-off processes, and the energy contours indicate the footstep strength and temporal changes of the footstep force. 
For energy features, we collectively include raw, smoothed, and logarithmic energy contours in our feature set. Based on empirical data analysis, using all three types of energy contours is superior to using any single energy contour.
The likely reason for this is that raw energy reveals fine details, smoothed energy reveals overall trends, and logarithmic energy highlights changes associated with perception. To this end, combining these three types of energy features makes the analysis more robust and not limited by any single representation. 

\subsubsection{Vibration-Related Features} 
We extract vibration-related features that characterize structural vibration signals correlated to footstep excitations. The vibration-related features are categorized into statistical features, time-domain features, frequency-domain features, time-frequency domain features, and compact signal representation features. 

Statistical features summarize the statistical properties of the overall amplitude distribution of the signal. Statistical features include the mean, median, standard deviation, maximum, range, skewness, kurtosis, number of peaks, number of valleys, autocorrelation, the slope of signal value increases or decreases in vibration signals, and spectral shape descriptors (spectrum/delta centroid, spectrum/delta crest, spectrum/delta decrease, spectrum/delta entropy, spectrum/delta flatness, spectrum/delta flux, spectrum/delta kurtosis, spectrum/delta skewness, spectrum/delta roll-off point, and spectrum/delta slope).
Metrics such as mean, median, standard deviation, and range provide insights into the signal’s overall shape and variability~\cite{kaur2018descriptive}, offering a foundation for understanding general footstep properties. For instance, more forceful and inconsistent footsteps result in higher standard deviation, range, and maximum values. The number of peaks and valleys reflects the consistency of footstep forces, while slope features indicate the rate of vibration amplitude changes. Autocorrelation reveals the periodicity of the signal, which is essential for recognizing patterns and regularities, reflecting deliberate pacing or hesitation in walking. Steeper slopes correspond to harder, sharper foot-floor contacts, whereas more gradual slopes represent softer or rolling footsteps. 
Spectral shape descriptors indicate the overall statistical properties of energy distribution in the frequency spectrum and its temporal changes. For example, the spectral/delta centroid represents the weighted average frequency and energy-concentrated frequency band~\cite{peeters2004large}, with higher values reflecting sharper footstep impacts. Spectral/delta entropy quantifies randomness in energy distribution~\cite{misra2004spectral, pikrakis2008speech}. Uneven or irregular footsteps exhibit higher entropy, while consistent, rhythmic steps correspond to lower values. 

Time domain features describe the behavior of vibration signals over time, corresponding to the footsteps' temporal dynamics, including jitter, shimmer, jitter relative average perturbation (jitter rap), and zero crossing rate. Jitter, shimmer, and jitter rap quantify the variability in signal frequency and amplitude~\cite{farrus2007jitter} which can show footstep stability and force duration, similar to its ability to detect voice anomalies~\cite{li2021acoustic}. The zero crossing rate is commonly used to distinguish between voiced and unvoiced speech~\cite{bachu2008separation} and can also help distinguish between smooth and erratic walking patterns.

Frequency domain features describe the distribution of signal energy across frequencies and the shape of the frequency spectrum. They represent structural responses to footstep force excitations. Frequency domain features include Fourier transform coefficients, harmonic ratios, and cepstral features. The Fourier transform coefficients capture the distribution of vibration energy over each frequency band, which helps identify the specific structural vibration mode patterns excited by footstep forces~\cite{fagert2021structure}. The harmonic ratio indicates the degree of harmonics in the signal with a larger value for a more consistent and stable gait pattern~\cite{bellanca2013harmonic}. Cepstral features reflect the rate of change in spectral components~\cite{gauer2021analysis}, which helps to identify the damping characteristics of structural vibration associated with foot-floor contact patterns and is robust across different structures~\cite{fagert2022clean}. 

Time-frequency domain features capture both temporal and frequency information from vibrations, representing the coupling between the footstep force dynamics over time and the structural response patterns induced by the footstep forces. Time-frequency domain features include wavelet spectra, Hilbert-Huang transform spectra, and fundamental frequency contours. The wavelet spectra show amplitude changes in the frequency components over time~\cite{torrence1998practical}, which is useful for distinguishing footstep events with similar frequencies at different gait phases. The Hilbert-Huang transform spectrum captures the signal envelope and phase information over time, making it effective for analyzing nonlinear and non-stationary vibration signals~\cite{huang2014hilbert, huang1998empirical}. This capability helps detect subtle footstep rhythm patterns. The fundamental frequency contour shows the variation of the dominant frequency over time. It reflects the excitation dynamics of the foot-floor interaction, revealing patterns such as sharp heel strike (a rising contour) or softer toe-off (a lowing contour).

Compact signal representation features encode signal structure and intrinsic patterns through data compression approaches, including linear prediction coefficients and Legendre polynomial coefficients. Linear prediction coefficients capture the temporal dependencies within a signal by modeling it as a linear combination of its past values~\cite{makhoul1975linear}, which helps to understand the underlying signal dynamics of footstep-induced floor vibrations. The Legendre polynomial coefficients represent information efficiently with good compression, capturing the essence of the signal with fewer parameters~\cite{ananthi2021audio}.

\subsection{Module 3: General Emotion Recognition Modeling and Simplification}
\begin{figure}[!t] 
  \centering
  \includegraphics[width=\linewidth]{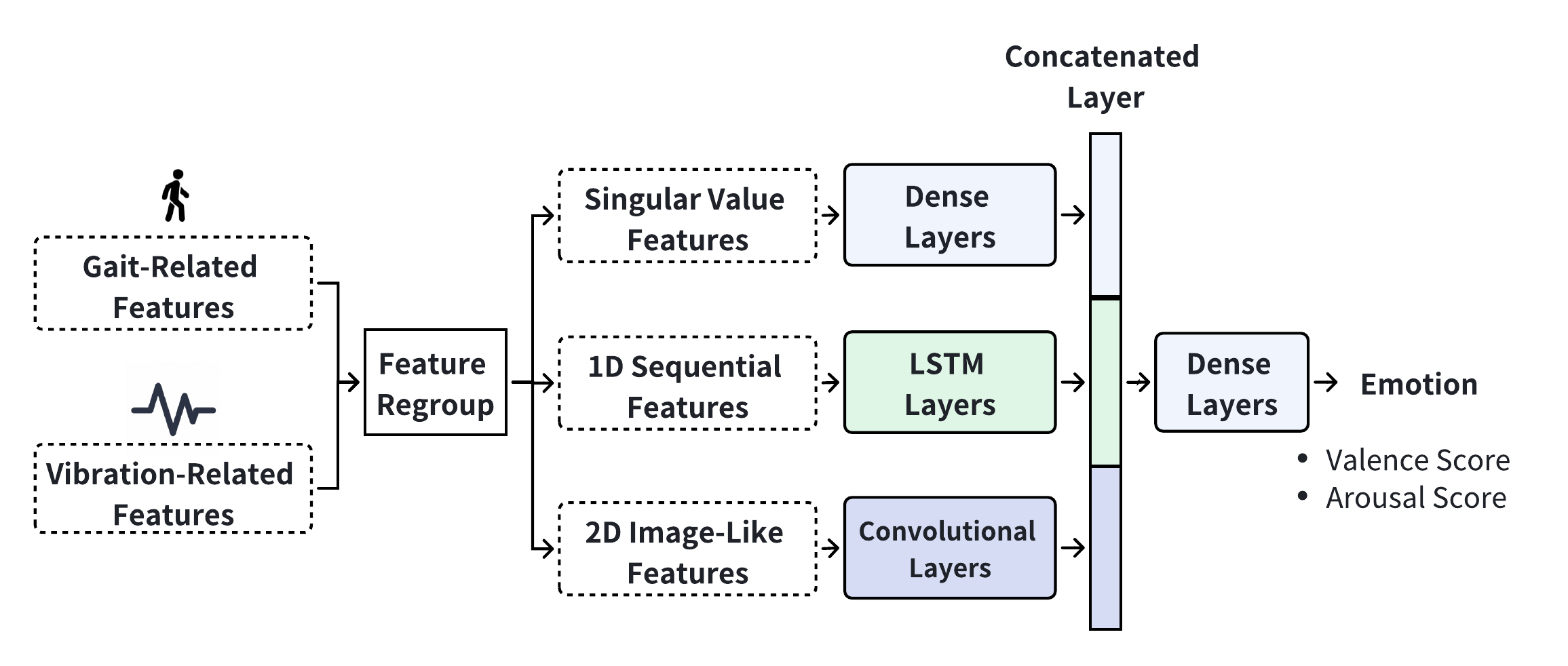}
  \caption{General Emotion Recognition Neural Network Architecture.}
  \label{fig:iteration_prune_2}
\end{figure}

We develop a general emotion recognition model to build a preliminary relationship between emotions and footstep-induced floor vibrations before inputting specific information about the target person’s footstep-induced floor vibrations. 
The model architecture is illustrated in Fig.~\ref{fig:iteration_prune_2}. The extracted emotion-sensitive features are regrouped into three types according to their data format: singular value features, one-dimensional sequential features, and two-dimensional image-like features. Distinct types of neural network layers are employed based on how each feature’s data format influences information representation and extraction. For singular value features, we use fully connected layers to decode information because these layers contain separate parameters for each input feature value, enabling direct modeling between the feature and the output. Sequential features have dependencies across time steps. So, they are processed through long short-term memory (LSTM) layers, which excel at capturing the temporal dependencies and deciding which data to retain or omit over sequences via the gating mechanism~\cite{hochreiter1997long}. Two-dimensional time-frequency spectrum image features, such as those derived from continuous wavelet transforms (CWT), are analyzed with convolutional layers. The convolutional layers are chosen because they can preserve the spatial and temporal relationships within the time-frequency spectrum image features. The output from each processing layer is combined and input into a multilayer perceptron with two fully connected layers, which outputs valence and arousal scores for the emotion score estimation task. In addition, dropout layers are incorporated after concatenation to reduce the overfitting caused by the noise and fluctuations within the training data~\cite{srivastava2014dropout}. The mean absolute error (MAE) is chosen as the loss function for the emotion score estimation due to its linear quantification of prediction errors, making it well-suited for regression tasks~\cite{willmott2005advantages}.

However, this emotion recognition model is easily overfitted with the large number of features and the limited size of the dataset due to the high cost of human experiments. To address this problem, we simplify the emotion recognition model by cutting down the least important parameters within the model through iterative pruning. The original feature set spans over 10,000 dimensions, suggesting a requirement for a dataset size at least ten times this number, or over 100,000 footstep samples, to train the network, as indicated by ~\cite{alwosheel2018your,baum1988size}. This requires 40 hours of continuous walking, not including the time allocated for emotion elicitation. The limited dataset can lead to model overfitting. On the other hand, conducting human experiments to gather such a vast dataset is prohibitively expensive and logistically challenging. 

To this end, we simplify the model using the iterative pruning method which iteratively cuts down the least important parameters from the neural network during training. The iterative pruning method has been proven effective in reducing the computational time while keeping the model performance ~\cite{ge2020compressed}. In this work, iterative pruning is used for model simplification to mitigate overfitting which improves the model performance in the situation with a limited dataset.
 The iterative pruning algorithm includes three phases: The initial phase involves training the network to obtain baseline parameter values without any reduction, providing a foundation for evaluating parameter importance. Subsequently, the least important parameters are pruned after each training epoch, systematically eliminating non-essential parameters until this phase concludes. Finally, pruning stops and the network is further trained with the pruned architecture to improve performance. The number of epochs of each phase and the pruning rate are hyperparameters selected through the grid search method. After iterative pruning, we obtain a general emotion recognition model that can estimate emotion from the input features extracted from the footstep-induced floor vibrations.
 
\subsection{Module 4: Personalized Emotion Recognition Modeling}

To address the large between-person variations among single footstep signals, the emotion recognition model is personalized with the incorporated footstep-induced floor vibrations from the target person (the individual whose emotions we aim to recognize). The personalization process contains two steps. First, we assess the gait similarity between the target person and the people in the training dataset and quantify the similarity using the gait similarity indices. Subsequently, we personalize the emotion recognition model by fine-tuning with these gait similarity indices as weights assigned to the training data in the loss function. During the fine-tuning, the model multiplies the loss by the sample’s weight instead of treating each sample’s loss equally. This allows the samples with higher weights to contribute more to the fine-tuning process. As a result, the model parameter updates prioritize minimizing loss for samples with higher weights, which represent samples that are more similar to the gait samples of the target person. 

We first compare gait similarity (quantified as gait similarity indices) between the target person and people in the training set based on their footstep-induced floor vibrations. 
The steps for computing gait similarity indices are detailed as follows:

\textbf{1) Sample-wise Similarity Calculation:} We first calculate the similarity between all sample pairs comprising one footstep sample from the target person ($P_T$) and the other from the training person ($P_i$, where i = 1, 2, ..., n) through Euclidean distance. The distances are calculated using the samples' embedded features corresponding to the output values of the concatenated layer obtained in the general emotion recognition model. We assume that these concatenated layers effectively represent essential gait patterns, as they integrate information across all feature types. The Euclidean distance between embedded features of the $k_1$-th sample ($E_T^{k_1}$) from the target person and  embedded features of the $k_2$-th sample ($E_i^{k_2}$) from training person $P_i$ is calculated as:
$$d(E_T^{k_1}, E_i^{k_2}) = ||E_T^{k_1} - E_i^{k_2}||_2,$$
where $||\cdot||_2$ is the L2 norm representing the dissimilarity of two embedding vectors~\cite{dokmanic2015euclidean}. The Euclidean distances indicate the inverse of the similarity between pairs of footstep samples, with shorter distances corresponding to greater similarity.

\textbf{2) Person-wise Similarity Calculation:} Considering the intra-person gait variability, for each training person, we average the Euclidean distances for all samples from the target and the training person to reduce the impact of any outlier footsteps or noise, leading to a more reliable and stable similarity representation. This averaging approach mitigates the potential distortion caused by outlier samples, ensuring a more stable weight normalization process. The average Euclidean distance ($D_{i}$) is calculated as follows: 
\[D_{i} = \frac{1}{K_1 \times K_2}\sum_{k_1=1}^{K_1}\sum_{k_2=1}^{K_2} d(E_T^{k_1}, E_i^{k_2}).\]
In this formula, $K_1$ denotes the number of gait samples from the target person ($P_T$), and $K_2$ represents the number of training people, $P_i$. The average Euclidean distances ($D_{i}$) indicate the overall similarity of the target person's ($P_T$) gait patterns and the training person's ($P_i$) gait patterns, with smaller $D_i$ values corresponding to greater similarity.

\textbf{3) Gait Similarity Normalization:} 
In order to quantify the gait similarities, the Gait Similarity Index (GSI) is established as the normalized inverse of this average distance ($D_i$):
\[
GSI_i = \frac{\widehat{GSI_i}}{\max_j(\widehat{GSI_j})}, \quad \text{where} \quad \widehat{GSI_i} = \frac{1}{D_{i}}.\]
For normalization, we divide each $\widehat{GSI}$ by the maximum $\widehat{GSI}$ value observed across the training people, constraining similarity indices to a range between 0 and 1, thereby facilitating weight assignment. Similar gait patterns usually have smaller Euclidean distances, thus corresponding to larger GSI values.

After assessing the gait similarity indices (GSI) between the target person and the individuals in the training set, we personalize the general emotion recognition model through fine-tuning. During this process, the parameters from the pruned general model are utilized as the initial model settings. Training samples are weighted by the training people's GSI values. Consequently, data from individuals with gait patterns similar to the target person receive higher weights, thus improving model performance for the target person by concentrating training on samples of similar gait patterns.
\section{Real-World Human Walking Evaluation}
\label{sec:eva}
To evaluate the performance of EmotionVibe, we conducted real-world walking experiments, collecting a dataset from 30 participants in the laboratory setting. 
Data from 20 participants were used for system evaluation, with data selection criteria detailed in Section~\ref{sec:explain}.
EmotionVibe achieved mean absolute errors of 1.11 for valence score estimation and 1.07 for arousal score estimation, within a score range of 1 to 9, representing 19.0$\%$ and 25.7 $\%$ error reductions respectively compared to the baseline method, which uses only gait-related features without personalization. 

\subsection{Human Walking Experiment Setup}

\begin{figure}[!t]
  \centering\includegraphics[width=\linewidth]{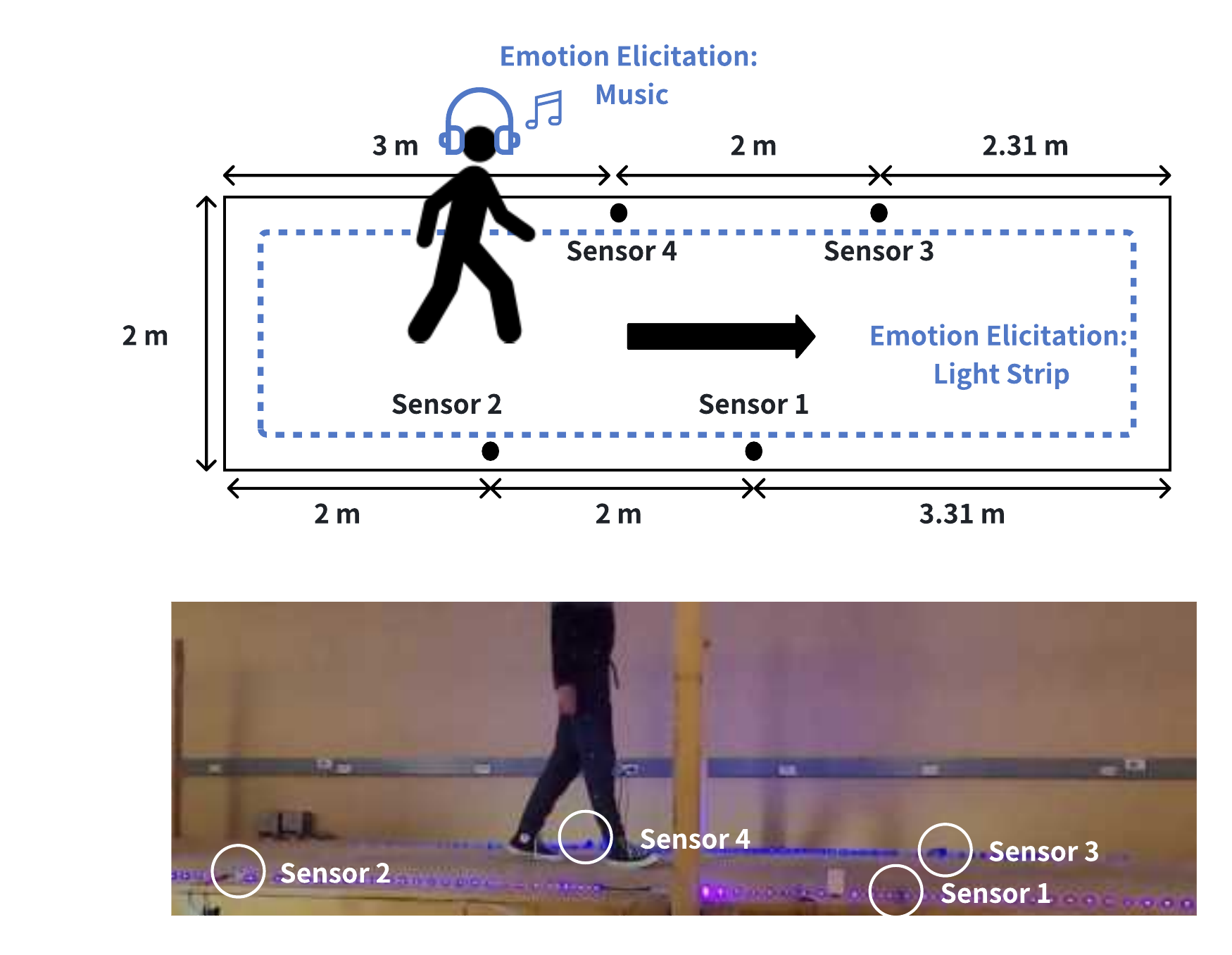}
  \caption{Real-World Walking Experiment Setup. }
  \label{fig:experiment}
\end{figure}

We conducted the real-world walking experiment on a wooden platform in our lab at Stanford University. The wooden platform, with dimensions of 7.31 meters in length and 2 meters in width, was equipped with four SM-24 geophone sensors~\cite{sm24_geophone_2006} attached to the edge of the floor using wax and glue (see Fig.~\ref{fig:experiment}). Following the Nyquist Sampling Theorem~\cite{6773024}, the sampling rate was set at 500 Hz, considering that the structural natural frequency is below 200 Hz.  All experiments were carried out in compliance with approved Institutional Review Board (IRB) protocols (Stanford IRB Protocol: 54912).

During the experiment, we used a variety of music clips and light stimuli, which have been proven effective methods for emotion elicitations, as detailed in Table~\ref{tab:emotion_stimulus}. Eight pairs of emotion elicitation were designed to elicitate four target emotional states: high valence high arousal, high valence low arousal, low valence high arousal, and low valence low arousal. The selected music clips were obtained from the PUMS database because of their efficacy in influencing human emotions~\cite{warrenburg2021pums}. Light stimuli included a shining light for high arousal scenarios, bright white for high valence low arousal, colorful shining colors for high valence high arousal, red and yellow for low valence high arousal, and dark blue for low valence low arousal situations. The effectiveness of light stimuli in eliciting emotions is supported by~\cite{wilson1966arousal,elliot2015color,valdez1994effects}. 
The Govee RGB LED Strip was used to provide varying lighting conditions while music clips were played through AirPods.

\begin{table}
\centering
\begin{tblr}{
  cells = {c},
  cell{1}{1} = {c=2}{},
  cell{1}{3} = {r=2}{},
  cell{1}{4} = {r=2}{},
  cell{3}{1} = {r=2}{},
  cell{3}{2} = {r=2}{},
  cell{3}{4} = {r=2}{},  
  cell{5}{1} = {r=2}{},
  cell{5}{2} = {r=2}{},
  cell{5}{4} = {r=2}{},  
  cell{7}{1} = {r=2}{},
  cell{7}{2} = {r=2}{},
  cell{7}{4} = {r=2}{},  
  cell{9}{1} = {r=2}{},
  cell{9}{2} = {r=2}{},
  cell{9}{4} = {r=2}{},  
  vlines,
  hline{1,3,5,7,9,11} = {-}{},
  hline{2} = {1-2}{},
  hline{4,6,8,10} = {3-4}{},
}
Emotion Type &         & Music Clips                                                               & Light Type              \\
Valence      & Arousal &                                                                          &                         \\
High         & High    & {Peer Gynt Suite No. 1, \\Op. 46, Mvt 4, In the Hall \\of the Mountain King} & {Colorful, \\Shining}   \\
             &         & {Prelude and Fugue \\No. 15 BWV 860,\\~I. Prelude in G Major}                &                         \\
High         & Low     & {Nocturne No. 2, \\Op. 9 in E Flat Major}                                    & {White, \\Steady}       \\
             &         & Blue in Green                                                                &                         \\
Low          & High    & Dracula, Vampire Hunters                                                     & {Red/Yellow, \\Shining} \\
             &         & High-Wire Stunts                                                             &                         \\
Low          & Low     & Adagio in G Minor                                                            & {Dark Blue, \\Steady}   \\
             &         & {The Seven Last Words of \\Jesus Christ, Op 51, Mvt 3}                       &                        
\end{tblr}

\caption{Emotion Elicitation Set in Real-World Walking Experiment.}
\label{tab:emotion_stimulus}
\end{table}

The experimental procedure was as follows: 
\begin{enumerate}
    \item The participant was guided to walk back and forth on the platform for 2 to 3 minutes with their initial emotions.
    \item The participant was guided to complete the Self-Assessment Manikin (SAM) survey scale (see Appendix~\ref{sec:appendix})~\cite{bradley1994measuring}, rating their emotional valence and arousal on a scale from 1 to 9.
    \item The participant was exposed to a randomly selected emotional elicitation from the eight categories outlined in Table~\ref{tab:emotion_stimulus} for 1 minute.
    \item The participant was guided to continue walking on the platform for 2 to 3 minutes, with music and lighting corresponding to the presented emotional elicitation.
    \item The participant was guided to complete the SAM survey scale again to assess their emotional state during the walk and evaluate the effectiveness of the emotion elicitations (see Appendix~\ref{sec:appendix}).
    \item Steps 3 to 5 were repeated with alternative emotion elicitations. This was done 8 times, each with a different set of elicitations from the specified emotion categories.
\end{enumerate}

During the experiment, each participant completed nine walking trials, with each trial lasting 2 to 3 minutes. 
Finally, we got 37,001 footstep samples from the four sensors in the dataset for evaluation.

\subsection{Dataset Overview and Effectiveness Validation}
\begin{figure}[!t]
  \centering\includegraphics[width=\linewidth]{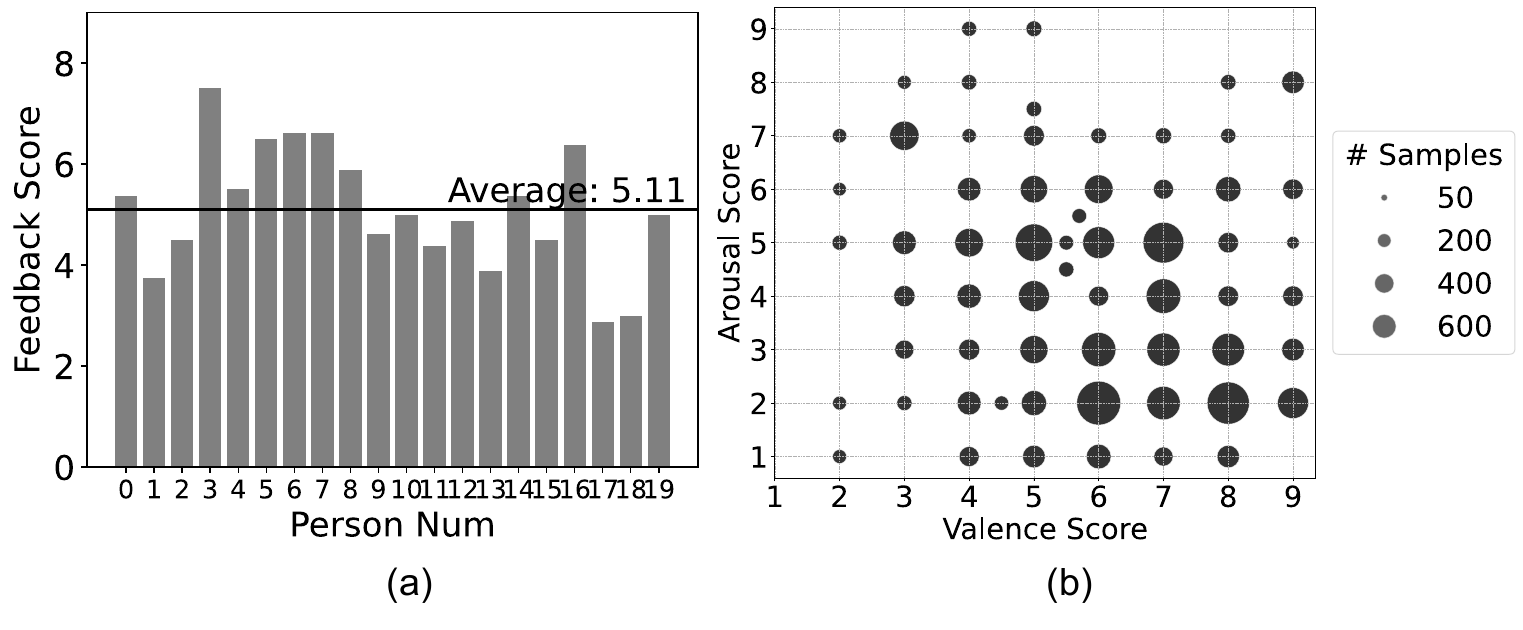}
  \caption{(a) The impact scores of emotional elicitation reported by participants, averaging 5.11, suggesting a moderate influence on their emotions. (b) Emotion scores distribution of the samples in our dataset, showing the emotion variability of our dataset. }
  \label{fig:feedback}
\end{figure}

We evaluated the effectiveness of emotion elicitation and the variability of emotions in the dataset through their feedback on emotion elicitation and the distribution of emotion scores in the survey results.
Each participant was asked to report an impact score from 1 to 9 for each set of emotional elicitation to assess the extent to which the emotional elicitation affected his or her emotion (bottom question in Appendix~\ref{sec:appendix}). 
The overall average of this impact score is 5.11, suggesting a moderately strong influence of our emotion elicitation on participants' emotions (see Fig.~\ref{fig:feedback} (a)). 
Fig.~\ref{fig:feedback} (b) shows the distribution of emotion scores within the 2D valence-arousal space. For valence, the measured values range from a minimum of 2 to a maximum of 9, while for arousal, the range spans from 1 to 9. This shows the wide variability in emotional states. 

\subsection{Hyper-parameter Selection}
\label{sec:hyper}
Hyperparameter selection for EmotionVibe was guided by a priori knowledge of human walking patterns, empirical data observations, and a grid search for optimal neural network hyperparameters. For single footstep signal segmentation in Module 1, we identified the highest peaks of wavelet transform coefficients within the 30–70 Hz range, corresponding to the fundamental frequency band of the wooden structure used in our experiment. The footstep signal was extracted as a 0.35 s segment spanning -0.15 s to 0.2 s around each peak. To calculate the full width at half maximum (FWHM), we analyzed the signal contour in the 100–250 Hz range for higher frequencies and 10–35 Hz for lower frequencies. Double support time was determined by the time difference between peaks in the [100, 200] Hz and [20, 30] Hz bands. This is based on the observation that the higher frequency part ([100, 200] Hz) mainly represents heel strike (double support initiation), and the lower frequency part ([20, 30] Hz) represents toe-off (double support termination). This is slightly narrower than in the previous calculation because when calculating the double support time, we want to reduce the noisy peaks and keep the main peak, so narrowing down the frequency band selection shows better performance. In the wavelet choosing part, the Morse wavelet was used. When calculating the energy contour of the signal, the window size was set to 0.05 s with a smooth span of 0.5 s for the smoothed energy.
The neural network architecture and hyperparameter selection were optimized via grid search. The convolutional block to process the 2D image-like features consisted of four convolutional layers, each followed by an average pooling layer with dropout regularization  (dropout rate = 0.5). The 1D sequential features were processed using an LSTM block with four units. 

\subsection{Emotion Recognition Results}
\label{sec:eva_result}

\begin{figure}[!t]
  \centering
  \includegraphics[width=0.8\linewidth]{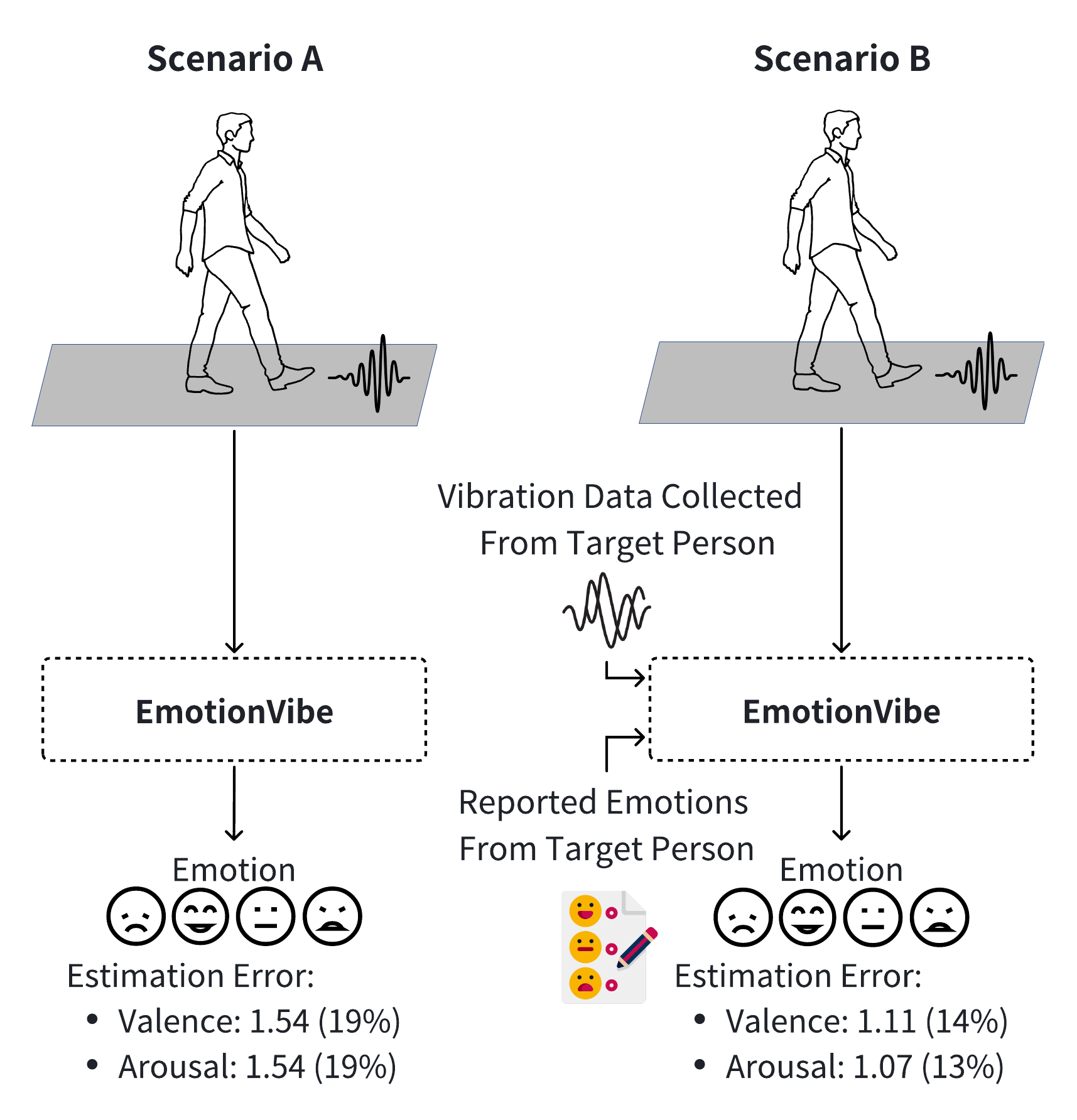}
  \caption{Evaluation scenarios: (a) Scenario A: EmotionVibe has no training data from the target person; (b) Scenario B: EmotionVibe is trained with 10 minutes of walking data from the target person.}
  \label{fig:usage}
\end{figure}

Our evaluation included two scenarios: Scenario A) none of the target person's data was used for model training and Scenario B) 10 minutes of walking data from the target person was used for training and the remaining part of the data (around 13 minutes of walking data) from the target person was used for testing (see Fig~\ref{fig:usage}). 
In both scenarios, we calculated the average emotion recognition results from 20 tests as the evaluation result, each involving a different individual as the target person.
In Scenario A, data from the 19 non-target participants were shuffled and split in a 9:1 ratio for training and validation. In Scenario B, the training and validation sets consisted of data from 19 non-target participants and 10-minute walk data from the target individuals, which were randomly assigned for training and validation in a 9:1 ratio. The test set consisted of approximately 13 additional minutes of walking data from the target individuals. The sets of tests, training, and validation were mutually exclusive. In addition, to prevent data leakage, we ensured that test data were not selected from the same walking trajectory (walking from one side of the platform to the other side once for about 10 seconds) as the training or validation set. 
The emotion recognition results for a trajectory were calculated based on the median scores of the emotion recognition results for each footstep in this trajectory.

\begin{figure}[!t]
  \centering
  \includegraphics[width=0.85\linewidth]{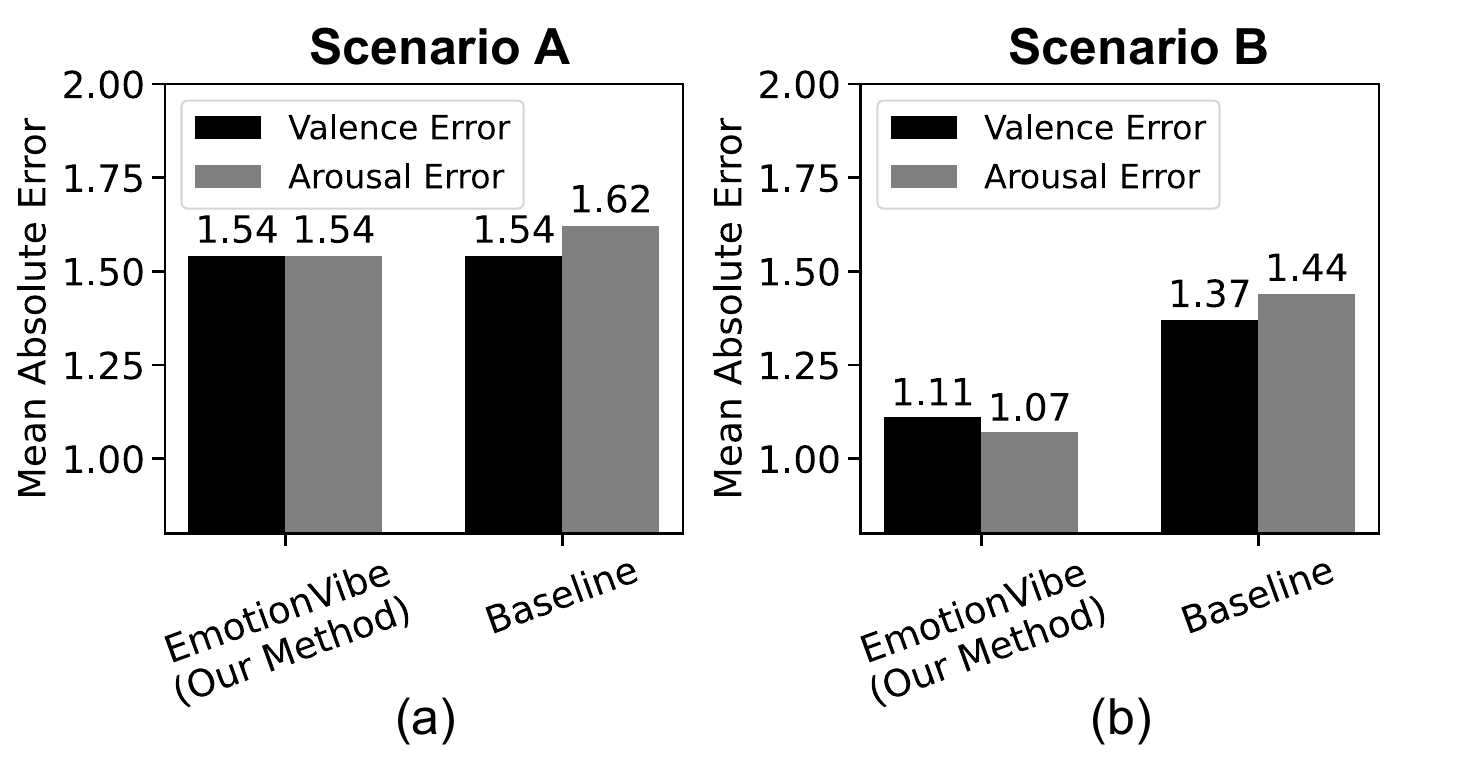}
  \caption{Emotion estimation error of EmotionVibe in (a) Scenario A (where none of the target person's data is known) and (b) Scenario B (where 10 minutes walking data from the target person is included in the training and validation). The baseline method relies solely on gait-related features and is modeled without personalization}
  \label{fig:result}
\end{figure}

\begin{figure}[t]
  \centering
  \includegraphics[width=0.95\linewidth]{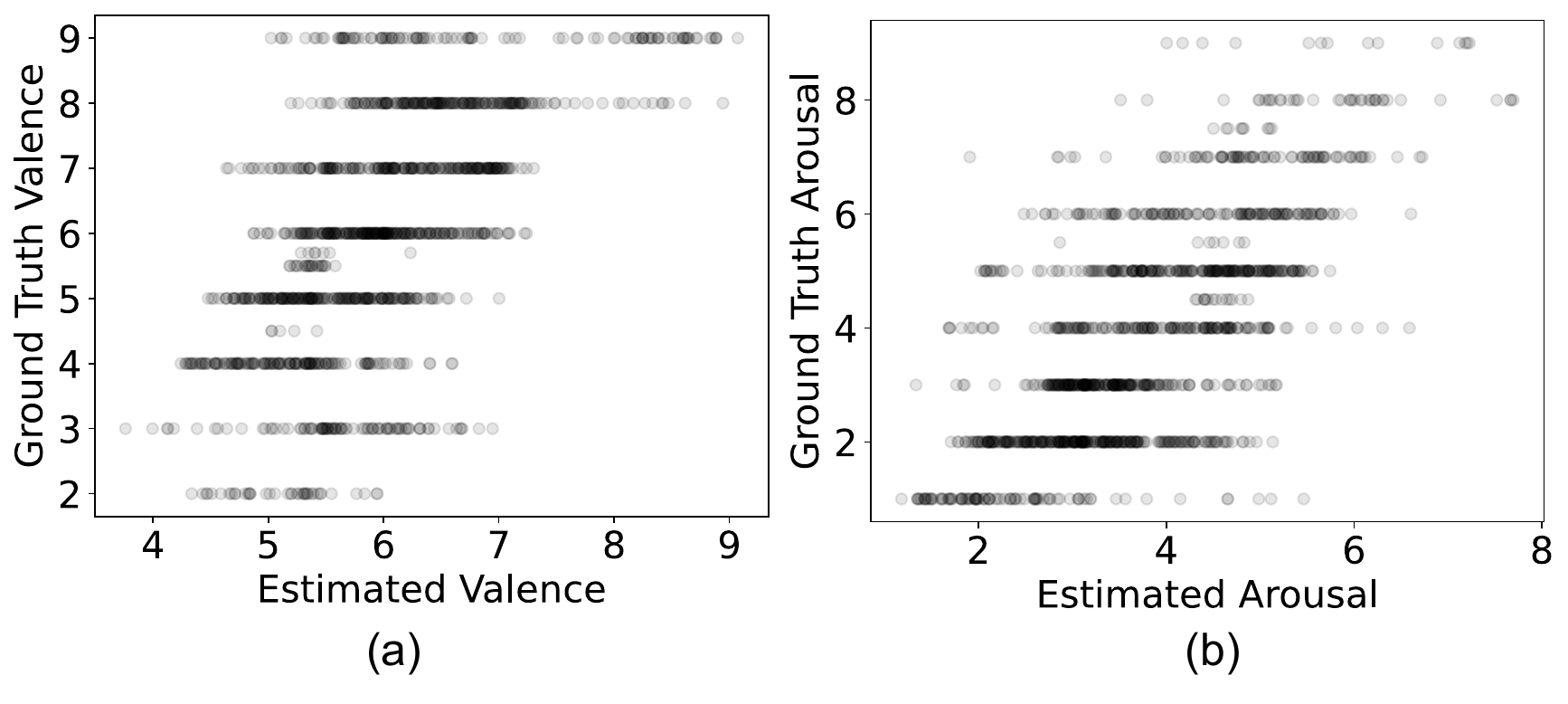}
  \caption{Scatterplots of the estimated and ground truth emotion scores for (a) valence estimation and (b) arousal estimation.}
  \label{fig:scatter}
\end{figure}

In both scenarios, EmotionVibe achieved accurate results for emotion score estimation.
The mean absolute errors (MAEs) are 1.54 (19.5$\%$ error rate) for both valence and arousal score estimation even without any data from the target person in the training data (Scenario A). When incorporating 90 walking trajectories (approximately 10 minutes of walking) from the target person into the training data, the MAEs are reduced to 
1.11 (13.9$\%$ error rate) and 1.07 (13.4$\%$ error rate) for valence and arousal score estimation, respectively, representing a 19.0$\%$ and 25.7$\%$ error reduction compared to the baseline method (see Fig.~\ref{fig:result}). The baseline model utilizes only gait-related features, which represent parameters derived from previous related works on gait patterns and emotions, without personalization. Fig.~\ref{fig:scatter} shows the plot comparing estimated scores against ground truth for valence and arousal in Scenario B, showing moderate linear correlations with the Pearson correlation score of 0.612 and 0.695, respectively. As a reference, the state-of-the-art method leveraging gait information from videos achieves 87.5$\%$ accuracy in emotion recognition~\cite{lima2024st}, which is comparable to our results but more intrusive and coarse-grained. EEG-based emotion recognition approaches reported an MAE of 0.48 (when scaled to the range of 1 to 9 scoring to match our scale) and a Pearson correlation of 0.8 between EEG-derived and self-reported emotion scores~\cite{galvao2021predicting}. However, EEG requires scalp electrodes, making it intrusive and potentially uncomfortable.

\section{Discussion}
\label{sec:discussion}

In this section, we discuss the effectiveness of each module in EmotionVibe through ablation tests, evaluate the robustness of EmotionVibe's performance, and justify the participant selection criteria for model training.

\subsection{Ablation Test}
\begin{figure}[!t]
  \centering
  \includegraphics[width=\linewidth]{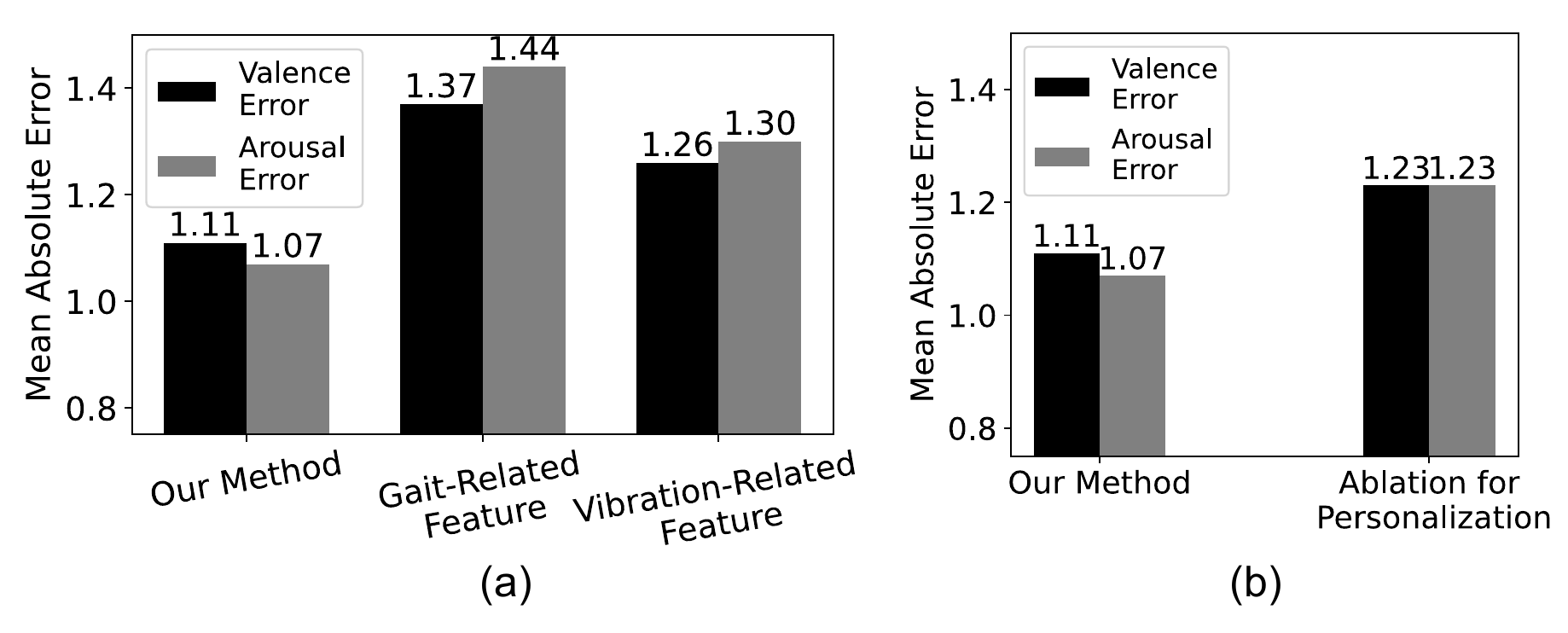}
  \caption{(a) Validating the Effectiveness of the Emotion-Sensitive Feature Set: Our method outperforms both methods that employ a single type of feature. (b) Validating the Effectiveness of Personalization: Without personalization, the emotion score estimation error increased by 12.8$\%$.}
  \label{fig:feature_ablation}
\end{figure}

We conducted ablation tests to demonstrate the effectiveness of emotion-sensitive feature sets and the personalization method. Emotion estimation error increased in both the case of removing part of the feature set and the case of removing the personalization procedure. To demonstrate the effectiveness of the emotion-sensitive feature set, we used Scenario B (defined in Section~\ref{sec:eva_result}) as an example. As shown in Fig.~\ref{fig:feature_ablation} (a), using any one of the two features (gait-related and vibration-related features) gave relatively good results but not as good as using a combination of these features. 
This finding indicates that both features are effective in capturing the relationship between human emotions and floor vibrations. Fig.~\ref{fig:feature_ablation} (b) shows the ablation results of the personalization method for Scenario B. Eliminating the personalization process and training the model with uniform weights across all individuals resulted in a 12.8$\%$ increase in estimation error, proving the effectiveness of personalization in improving emotion recognition performance.

\subsection{System Robustness Test}
In this subsection, we evaluated the robustness of EmotionVibe's performance across individuals and its performance improvement as it progressively collected footstep samples with reported emotions from the target person.

\subsubsection{Robustness of System Performance Across Individuals}

\begin{figure}[!t]
  \centering
  \includegraphics[width=0.9\linewidth]{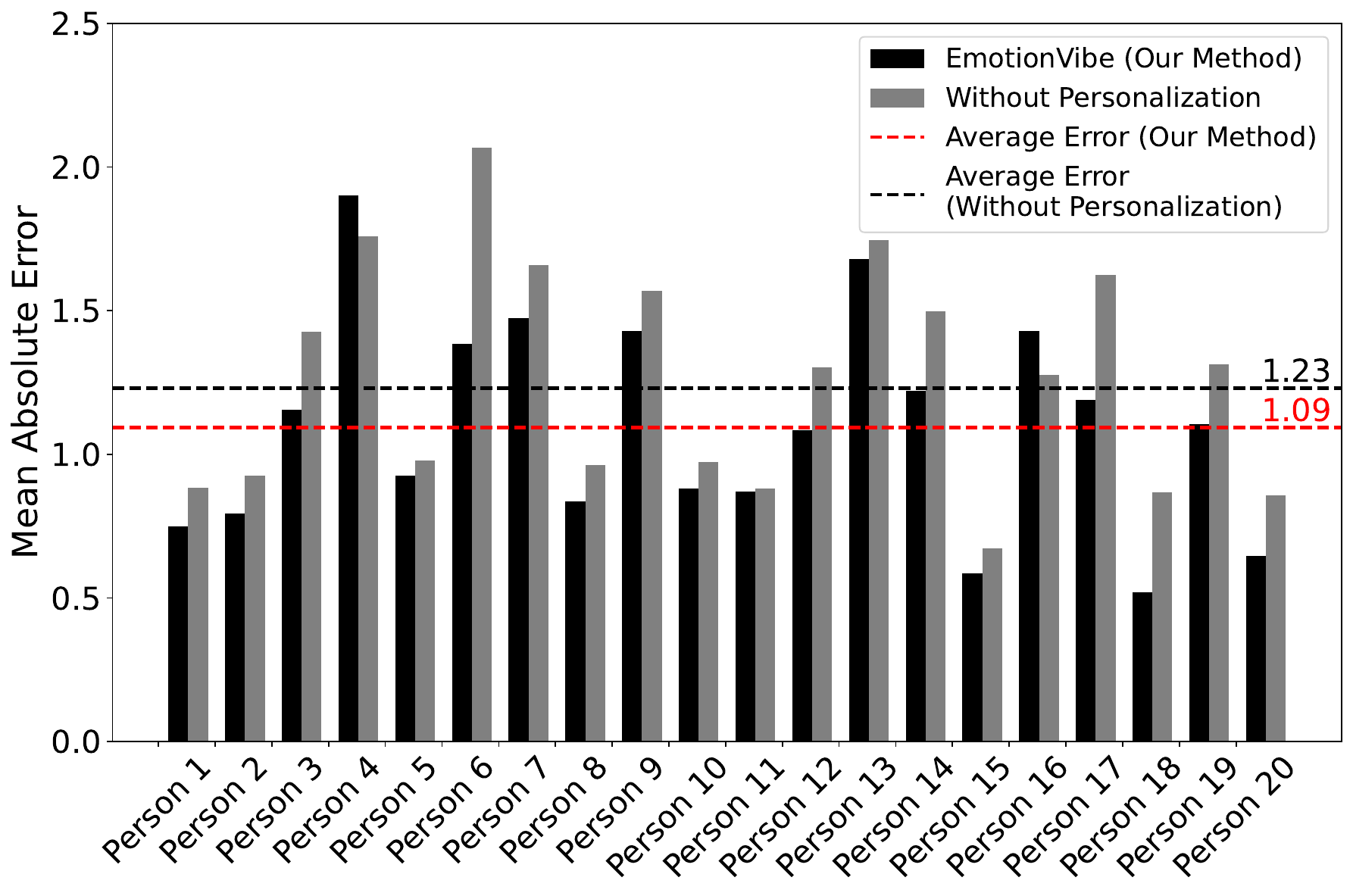}
  \caption{Mean Absolute Error (MAE) of EmotionVibe for 20 participants (compared to the method without personalization). EmotionVibe shows lower errors on the emotion estimation task for most participants.}
  \label{fig:personcompare}
\end{figure}

We evaluated the robustness of our system performance across different individuals and observed a consistent error reduction through personalization for most participants (see Fig.~\ref{fig:personcompare}). EmotionVibe achieved the best performance (lowest estimation error) for Person 18, with an MAE of 0.52 (valence score estimation error: 0.48, arousal score estimation error: 0.56).
In contrast, the model performed worst (highest estimation error) on Person 4's data, with an MAE of 1.9 (valence score estimation error: 2.18, arousal score estimation error: 1.68). 

Based on our observation, the variance in the test accuracy of emotion estimation between individuals is influenced by two key factors: the intrinsic variability of the target person's emotions and the general similarity in gait between the target person and those in the training set. Generally, a higher variance in emotional states in test data leads to a larger estimation error for the target person, as increased fluctuation in emotions makes it more challenging for the model to learn stable patterns. Additionally, when an individual's gait pattern is less similar to those in the training set, the model struggles to generalize effectively, resulting in a higher estimation error.
Person 18’s emotional variability is relatively low, as indicated by a low variance of 0.25 for the valence score and 0.48 for the arousal score. A smaller emotional range allows the model to learn more stable patterns, reducing estimation errors. Additionally, Person 18’s gait similarity to other individuals in the training set is higher than the average, enabling the model to generalize more effectively based on learned gait-emotion relationships.
In contrast, Person 4 exhibits a highly variable emotional range, with valence and arousal score variances of 7.07 and 9.16, respectively. This wide distribution increases the difficulty of accurately mapping gait features to emotional states. In addition, Person 4's gait pattern is distinctively different from the majority of individuals in the training set, as indicated by a median gait similarity that is 11.3$\%$ lower than the overall average before normalization. The reduced similarity limits the model’s ability to generalize effectively, leading to a higher estimation error.

Notably, even in the worst case, the MAE of our method remained below 2 within a range of 1 to 9, indicating a consistently high level of performance in emotion estimation tasks. Furthermore, the system exhibited significant error reduction for most people, with the most notable improvement observed for Person 6, where the MAE decreased from 2.07 to 1.39. However, the system did not reduce the error for Persons 4 and 16, potentially due to their unique gait patterns, which lack similar samples in the training dataset. Specifically, we observed that Person 4 wore boots with a hard heel, and Person 16 was the only participant wearing high heels, distinguishing their gait patterns from the rest of the participants. These outlier gait patterns make it challenging for the model to accurately estimate emotions without similar samples in the training dataset for reference. A potential solution could involve enriching the dataset with more diverse data, including examples of gait patterns of high heels and hard-heeled boots.

\subsubsection{System Performance with Varying Target Person Samples}
\label{sec:sys_cha_2}
\begin{figure}[!t]
  \centering
  \includegraphics[width=0.75\linewidth]{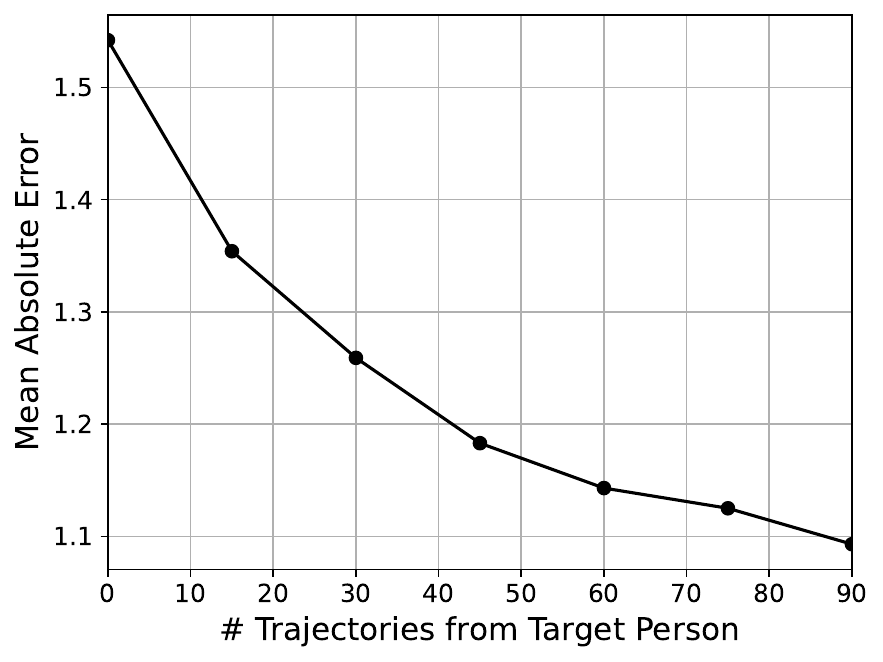}
  \caption{The performance of EmotionVibe improved as more data was collected from the target individual, leading to a decrease in estimation error.}
  \label{fig:fewshot}
\end{figure}

The performance of EmotionVibe gradually improves when footstep samples are obtained from the target person. We noted a decrease in the emotion estimation error as the quantity of input data from the target person increased (see Fig.~\ref{fig:fewshot}). When no footstep sample from the target person was incorporated into the training set, EmotionVibe achieved mean absolute errors (MAE) of 1.54 and 1.54 for valence and arousal estimation, respectively. When adding about 90 walking trajectories of the target person to the training set, the system performance improved, with reduced MAEs of 1.11 and 1.07 for valence and arousal score estimation, respectively. 

\subsection{Explanation for Participant Selection Criteria}
\label{sec:explain}
To ensure the reliability and validity of our analysis, we selected 20 participants from 30 participants for training data, excluding those who: (1) kicked the sensors while walking, (2) misunderstood the valence-arousal scale, leading to inconsistent or unreliable emotion reports, (3) left early without completing the experiment, and (4) exhibited small emotional response to elicitation, reporting an average impact score of emotional elicitation (Bottom question in Appendix~\ref{sec:appendix}) of less than 2 out of 9. The first three cases introduce erroneous or incomplete data.
Data from participants who exhibited small emotional responses to emotional elicitation is excluded because their low emotional variance can lead to model bias during training. 
Ideally, the model should learn the intended relationship:
``Changes in gait reflect changes in emotion."
However, when certain participants display little to no emotional variation, the model risks learning an unintended shortcut:
``If I recognize this person’s footsteps, I can predict their emotion as always being the same."
While this shortcut may reduce training error, it does not accurately capture the relationship between gait and emotional changes, thereby undermining the model’s validity.
Although including these data in the training set may introduce bias, EmotionVibe still handles them well when they are in the test set. EmotionVibe resulted in a mean absolute error on the test set from these participants for valence and arousal estimations of approximately 0.5, corresponding to an error rate of around 6.25$\%$.

\section{Conclusions and Future Work}
\label{sec:conclusion}
This paper introduces EmotionVibe, a novel emotion recognition system using footstep-induced floor vibrations. The physical insight of our system is that human emotions affect their gait patterns, which in turn influence the footstep-induced floor vibrations. Our main innovation lies in two aspects. Firstly, we develop two emotion-sensitive features, including gait-related and vibration-related features to capture the complex and indirect relationship between emotion and floor vibrations. Secondly, we personalize the emotion recognition system by assigning higher weights to people with similar gait patterns to the target person. To evaluate the effectiveness of EmotionVibe, we conducted a real-world walking experiment involving 20 participants, with a dataset of 37,001 footstep samples. Our system achieved mean absolute errors of 1.11 and 1.07 for valence and arousal score estimation respectively, achieving 19.0$\%$ and 25.7$\%$ error reductions compared to the baseline method. Our study provides a non-intrusive and privacy-friendly emotion recognition system that expands the possibilities of deploying emotion recognition systems in smart home environments for mental health monitoring and emotion-based recommendations.

In the future, we aim to extend EmotionVibe to real-world applications to recognize mixed emotions, adapt to more diverse structural environments, and take various types of human activities into account:

\textit{Expanding to Include Mixed Emotions:}
Human emotions are often mixed, with individuals experiencing multiple emotions simultaneously~\cite{larsen2014case,williams2002can,larsen2011further}. These mixed emotions can lead to subtle and nuanced variations in gait and other behaviors, making it challenging to categorize them into distinct emotional states. Future research could explore techniques such as dynamic pattern recognition to better identify and interpret these mixed emotional states.

\textit{Expanding to Diverse Structures:}
In the future, we aim to adapt EmotionVibe to various architectural structures. This requires the consideration of unique acoustics and vibration characteristics of different building materials such as wood, concrete, and steel. Developing a structure-invariant system can broaden its practical use in real-world environments.

\textit{Expanding to Detect Various Activities:}
Emotions are reflected in various daily activities, including speaking, gaming, and typing. Each type of activity corresponds to a unique behavioral pattern and physical interaction with the structure. Future research will explore emotion recognition using structural vibrations induced by various activities.

\section*{Acknowledgments}
This research is supported by the Stanford CEE-PhD Fellowship and Stanford Blume Fellowship at Stanford.

{\appendix[Emotion Survey]
\label{sec:appendix}

\begin{figure}[H]
  \centering
  \includegraphics[width=\linewidth]{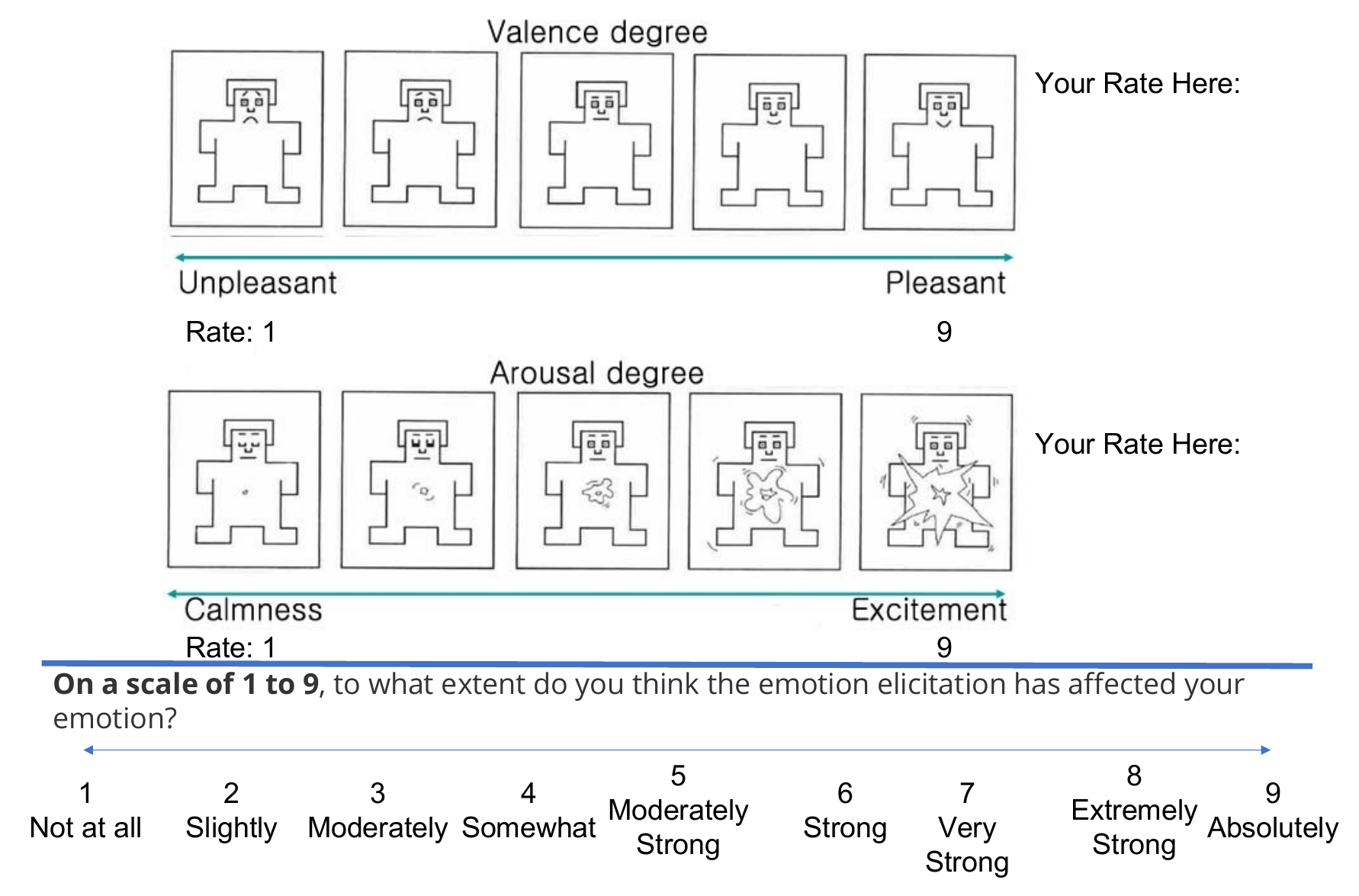}
  \caption{Emotion Survey Scale for Real-World Human Walking Experiments}
  \label{fig:survey}
\end{figure}}


 
%

\bibliographystyle{IEEEtran}
\bibliography{yuyan_v2}  

\vfill

\end{document}